\documentclass[journal,draftcls,onecolumn,11pt]{IEEEtran}
\usepackage{graphicx,psfrag,amsmath,epsfig,amssymb,epsf,cite,latexsym,mathrsfs,subfigure,url,array,dsfont}

\newcommand{\dotle} {\mbox{$\:\stackrel{\centerdot}{\le}\:$}}

\newcommand{\calA}{{\cal A}}
\newcommand{\calB}{{\cal B}}
\newcommand{\calC}{{\cal C}}

\newcommand{\calE}{{\cal E}}

\newcommand{\calI}{{\cal I}}
\newcommand{\calK}{{\cal K}}
\newcommand{\calM}{{\cal M}}

\newcommand{\calS}{{\cal S}}

\newcommand{\calU}{{\cal U}}
\newcommand{\calV}{{\cal V}}
\newcommand{\calW}{{\cal W}}
\newcommand{\calX}{{\cal X}}
\newcommand{\calY}{{\cal Y}}
\newcommand{\calZ}{{\cal Z}}

\newcommand{\scrP}{\mathscr{P}}
\newcommand{\scrW}{\mathscr{W}}

\newcommand{\bs}{{\mathbf s}}
\newcommand{\bS}{{\mathbf S}}
\newcommand{\bu}{{\mathbf u}}
\newcommand{\bU}{{\mathbf U}}

\newcommand{\bw}{{\mathbf w}}
\newcommand{\bW}{{\mathbf W}}
\newcommand{\bx}{{\mathbf x}}
\newcommand{\bX}{{\mathbf X}}
\newcommand{\by}{{\mathbf y}}
\newcommand{\bY}{{\mathbf Y}}
\newcommand{\bz}{{\mathbf z}}

\newcommand{\tp}{\tilde{p}}

\newcommand{\tby}{\tilde{\mathbf y}}

\def\eE{{\Bbb E}}

\def\rR{{\Bbb R}}

\newcommand{\oI}{{\overset{\circ}{I}}}

\newtheorem{definition}{Definition}[section]

\newtheorem{theorem}{Theorem}[section]
\newtheorem{lemma}[theorem]{Lemma}
\newtheorem{proposition}[theorem]{Proposition}

\renewcommand{\theequation}{\arabic{section}.\arabic{equation}}
\begin{document}

\title{Blind Fingerprinting}

\author{Ying Wang and Pierre~Moulin 
\thanks{Y.~Wang is with Qualcomm, Bedminster, NJ.
P.~Moulin is with the ECE Department, the Coordinated Science Laboratory,
and the Beckman Institute at the University of Illinois at Urbana-Champaign, Urbana, IL 61801, USA.
Email: {\tt moulin@ifp.uiuc.edu}. This work was supported by NSF under grants CCR 03-25924,
CCF 06-35137 and CCF 07-29061. Part of this work was presented at ISIT'06 in Seattle, WA.}}

\maketitle

\begin{abstract}
We study blind fingerprinting, where the host sequence into which fingerprints
are embedded is partially or completely unknown to the decoder. This problem relates to a multiuser
version of the Gel'fand-Pinsker problem. The number of colluders and the collusion channel
are unknown, and the colluders and the fingerprint embedder are
subject to distortion constraints. 

We propose a conditionally constant-composition random binning scheme and a universal decoding
rule and derive the corresponding false-positive and false-negative error exponents.
The encoder is a stacked binning scheme and makes use of an auxiliary random sequence.
The decoder is a {\em maximum doubly-penalized mutual information decoder},
where the significance of each candidate coalition is assessed relative to
a threshold that trades off false-positive and false-negative error exponents.
The penalty is proportional to coalition size and is a function
of the conditional type of host sequence.
Positive exponents are obtained at all rates below a certain value,
which is therefore a lower bound on public fingerprinting capacity.
We conjecture that this value is the public fingerprinting capacity.
A simpler threshold decoder is also given, which has similar universality
properties but also lower achievable rates.
An upper bound on public fingerprinting capacity is also derived.
\end{abstract}


\newpage

\section{Introduction}

Content fingerprinting finds applications to document protection
for multimedia distribution, broadcasting, and traitor tracing
\cite{Boneh95,Cox97,Wu04,Liu06}. A covertext---image, video, audio, or
text---is to be distributed to many users. A fingerprint, a mark unique
to each user, is embedded into each copy of the covertext.
In a collusion attack, several users may combine
their copies in an attempt to ``remove'' their fingerprints and to
forge a pirated copy. The distortion between the pirated copy and
the colluding copies is bounded by a certain tolerance level. To
trace the forgery back to the coalition members, we need
fingerprinting codes that can reliably identify the fingerprints
of those members. Essentially, from a communication viewpoint, the
fingerprinting problem is a multiuser version of the watermarking
problem~\cite{Moulin02,Moulin03,Somekh05,Somekh07,Barg07,Moulin08}.
For watermarking, the attack is by one user and is based on
one single copy, whereas for fingerprinting, the attack is
modeled as a multiple-access channel (MAC).
The covertext plays the role of side information to the encoder
and possibly to the decoder.

Depending on the availability of the original covertext to the
decoder, there are two basic versions of the problem:
private and public. In the {\em private fingerprinting} setup, the
covertext is available to both the encoder and decoder. In the {\em
public fingerprinting} setup, the covertext is available to the
encoder but not to the decoder, and thus decoding performance
is generally worse. However public fingerprinting presents an important
advantage over private fingerprinting, in that it does not
require the vast storage and computational resources that are
needed for media registration in a large database. For example, a DVD
player could detect fingerprints from a movie disc and refuse to
play it if fingerprints other than the owner's are present. Or Web
crawling programs can be used to automatically search for
unauthorized content on the Internet or other public
networks~\cite{Wu04}. 

The scenario considered in this paper is one where a degraded version $S^d$
of each host symbol $S$ is available to the decoder.
Private and public fingerprinting are obtained as special cases with $S^d = S$
and $S^d = \emptyset$, respectively. We refer to this scenario as either
{\em blind} or {\em semiprivate fingerprinting}. The motivation is analogous
to semiprivate watermarking \cite{Moulin07}, where some information
about the host signal is provided to the receiver in order to improve decoding
performance. This may be necessary to guarantee an acceptable performance
level when the number of colluders is large.

The capacity and reliability limits of \emph{private fingerprinting}
have been studied in~\cite{Somekh05,Somekh07,Barg07,Moulin08}.
The decoder of \cite{Moulin08} is a variation of Liu and Hughes' minimum equivocation
decoder \cite{Liu96}, accounting for the presence of side information and
for the fact that the number of channel inputs is unknown.
Two basic types of decoders are of interest: detect-all and detect-one.
The \emph{detect-all} decoder aims to catch all members of the
coalition and an error occurs if some colluder escapes detection.
The \emph{detect-one} decoder is content with catching at
least one of the culprits and an error occurs only when none of
the colluders is identified. A third type of error (arguably the
most damaging one) is a {\em false positive}, by which the decoder
accuses an innocent user.

In the same way as fingerprinting is related to the MAC problem,
blind fingerprinting is related to a multiuser extension of the Gel'fand-Pinsker
problem. The capacity region for the latter problem is unknown.
An inner region, achievable using random binning, was given in \cite{Somekh04}.

This paper derives random-coding exponents and an upper bound on detect-all
capacity for semiprivate fingerprinting.
Neither the encoder nor the decoder know the number of colluders.
The collusion channel has arbitrary memory but is subject to a distortion constraint
between the pirated copy and the colluding copies. Our fingerprinting
scheme uses random binning because, unlike in
the private setup, the availability of side information to the
encoder and decoder is asymmetric. To optimize the error exponents, we propose
an extension of the \emph{stacked-binning} scheme that was developed
for single-user channel coding with side information~\cite{Moulin07}.
Here the codebook consists of a stack of variable-size codeword-arrays indexed
by the conditional type of the covertext sequence.
The decoder is a {\em minimum doubly-penalized equivocation} (M2PE) decoder or
equivalently, a {\em maximum doubly-penalized mutual information} (M2PMI) decoder. 

The proposed fingerprinting system is universal in that it can cope with unknown
collusion channels and unknown number of colluders, as in the private fingerprinting
setup of \cite{Moulin08}. A tunable parameter $\Delta$ trades off false-positive
and false-negative error exponents. The derivation of these exponents
combines techniques from \cite{Moulin08} and \cite{Moulin07}.
A preliminary version of our work, assuming a fixed number of colluders,
was given in \cite{Wang06,Wang06b}.

\subsection{Organization of This Paper}

A mathematical statement of our generic fingerprinting problem is given in
Sec.~\ref{sec:problem statement},
together with the basic definitions of error probabilities, capacity,
error exponents, and fair coalitions. 
Sec.~\ref{sec:encoder} presents our random coding scheme.
Sec.~\ref{sec:simple} presents a simple but suboptimal decoder that compares
empirical mutual information scores between received data and individual fingerprints,
and outputs a guilty decision whenever the score exceeds a certain tunable threshold.
Sec.~\ref{sec:joint} presents a joint decoder that assigns a penalized empirical mutual
information score to candidate coalitions and selects the coalition with 
the highest score. 
Sec.~\ref{sec:C-public} establishes an upper bound on blind fingerprinting
capacity under the detect-all criterion. Finally, conclusions are given in
Sec.~\ref{sec:conclusion}.
The proofs of the theorems are given in appendices.

\subsection{Notation}
\label{sec:notation}

We use uppercase letters for random variables, lowercase letters
for their individual values, calligraphic letters for finite alphabets,
and boldface letters for sequences. We denote by $\calM^\star$
the set of sequences of arbitrary length (including 0) whose elements
are in $\calM$.
The probability mass function (p.m.f.) of a random variable $X \in
\calX$ is denoted by $p_X=\{p_X(x),\,x \in \calX\}$. The entropy
of a random variable $X$ is denoted by $H(X)$, and the mutual
information between two random variables $X$ and $Y$ is denoted by
$I(X;Y)=H(X)-H(X|Y)$. Should the dependency on the underlying p.m.f.s
be explicit, we write the p.m.f.s as subscripts, e.g., $H_{p_X}(X)$
and $I_{p_X,p_{Y|X}}(X;Y)$. The Kullback-Leibler divergence
between two p.m.f.s $p$ and $q$ is denoted by $D(p||q)$; the
conditional Kullback-Leibler divergence of $p_{Y|X}$ and $q_{Y|X}$
given $p_X$ is denoted by
$D(p_{Y|X}||q_{Y|X}|p_X)=D(p_{Y|X}\,p_X||q_{Y|X}\,p_X)$.
All logarithms are in base 2 unless specified otherwise.

Denote by $p_\bx$ the type, or empirical p.m.f. induced by a sequence $\bx
\in \calX^N$. The type class $T_\bx$ is the set of all
sequences of type $p_\bx$. Likewise, we denote by $p_{\bx\by}$
the joint type of a pair of sequences $(\bx, \by) \in \calX^N \times
\calY^N$ and by $T_{\bx\by}$ the type class associated with
$p_{\bx\by}$. 
The conditional type $p_{\by|\bx}$ of a pair of sequences ($\bx, \by$)
is defined by $p_{\bx\by}(x,y)/p_{\bx}(x)$ for
all $x \in \calX$ such that $p_{\bx}(x) > 0$. The conditional type
class $T_{\by|\bx}$ given $\bx$, is the set of all sequences $\tilde{\by}$
such that $(\bx, \tilde{\by}) \in T_{\bx\by}$. We denote by $H(\bx)$ the empirical
entropy of the p.m.f. $p_{\bx}$, by $H(\by|\bx)$ the empirical conditional
entropy, and by $I(\bx;\by)$ the empirical
mutual information for the joint p.m.f. $p_{\bx\by}$. 


We use the calligraphic fonts $\scrP_X$ and $\scrP_X^{[N]}$ to represent the set
of all p.m.f.s and all empirical p.m.f.'s, respectively, on the alphabet
$\calX$. Likewise, $\scrP_{Y|X}$ and $\scrP_{Y|X}^{[N]}$ denote the
set of all conditional p.m.f.s and all empirical conditional p.m.f.'s on
the alphabet $\calY$. A special symbol $\scrW_K$ will be used
to denote the feasible set of collusion channels $p_{Y|X_1, \cdots, X_K}$
that can be selected by a size-$K$ coalition.

Mathematical expectation is denoted by the symbol $\eE$.
The shorthands $a_N \doteq b_N$ and $a_N \dotle b_N$
denote asymptotic relations in the exponential scale, respectively
$\lim_{N \to \infty}\frac{1}{N}\log \frac{a_N}{b_N}=0$ and
$\limsup_{N \to \infty}$ $\frac{1}{N}\log \frac{a_N}{b_N}\le 0$.
We define $|t|^+ \triangleq \max(t,0)$ and $\exp_2(t)\triangleq 2^t$.
The indicator function of a set $\calA$ is denoted by $\mathds 1 _{\{x \in \calA\}}$.
Finally, we adopt the convention that the minimum of a function
over an empty set is $+\infty$ and the maximum of a function over
an empty set is 0.

\section{Statement of the Problem}
\label{sec:problem statement}

\subsection{Overview}


\begin{figure}[hbt]
\begin{center}
\includegraphics[width=13cm]{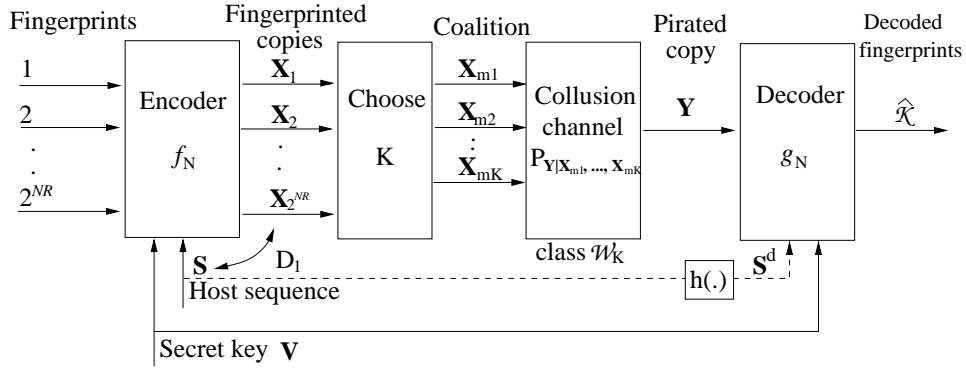}
\end{center}
\caption{Model for semiprivate (blind) fingerprinting game, where $\bS^d$ is a degraded
version of the covertext $\bS$. Private and public fingerprinting arise as special cases
with $\bS^d = \bS$ and $\bS^d = \emptyset$, respectively.}
\label{fig:Public FP}
\end{figure}

Our model for blind fingerprinting is diagrammed in Fig.~\ref{fig:Public FP}.
Let $\calS$, $\calX$, and $\calY$ be three finite alphabets.
The covertext sequence $\bS=(S_1, \cdots, S_N) \in \calS^N$ consists of
$N$ independent and identically distributed (i.i.d.) samples drawn
from a p.m.f. $p_S(s)$, $s \in \calS$. A random variable $V$ taking values 
in an alphabet $\calV_N$ is shared between encoder and decoder, and not publicly
revealed. The random variable $V$ is independent of $\bS$ and plays the role of
a cryptographic key.
There are $2^{NR}$ users, each of which receives a fingerprinted copy:
\begin{equation}
  \bX_m = f_N(\bS, V, m), \quad 1 \le m \le 2^{NR},
\end{equation}
where $f_N: \calS^N \times \calV_N \times \{ 1, \cdots, 2^{NR} \} \to \calX^N$
is the encoding function, and $m$ is the index of the user.
The encoder binds each fingerprinted copy $\bx_m$ to the covertext $\bs$
via a distortion constraint. Let $d~:~\calS \times \calX \to \rR^+$
be the distortion measure and $d^N(\bs,\bx) = \frac{1}{N} \sum_{i=1}^N d(s_i,x_i)$
the extension of this measure to length-$N$ sequences.
The code $f_N$ is subject to the distortion constraint
\begin{equation}
   d^N(\bs,\bx_m) \le D_1 \quad 1 \le m \le 2^{NR} .
\label{eq:D1}
\end{equation}

Let $\calK \triangleq \{m_1,\,m_2\,\cdots,\,m_K\}$ be a coalition of $K$ users,
called colluders. No constraints are imposed on the formation of coalitions.
The colluders combine their copies $\bX_{\calK} \triangleq \{\bX_m , \,m \in \calK\}$
to produce a pirated copy $\bY \in \calY^N$. Without loss of generality, we assume
that $\bY$ is generated stochastically as the output of a collusion channel
$p_{\bY|\bX_{\calK}}$. Fidelity constraints are
imposed on $p_{\bY|\bX_{\calK}}$ to ensure that $\bY$ is ``close''
to the fingerprinted copies $\bX_m , \,m \in \calK$.
These constraints can take the form of distortion constraints,
analogously to (\ref{eq:D1}). They are formulated below and result
in the definition of a feasible class $\scrW_K$ of attacks.

The decoder knows neither $K$ nor $p_{\bY|\bX_{\calK}}$ selected by the $K$ colluders
and has access to the pirated copy $\bY$, the secret key $V$, as well as to $\bS^d$,
a degraded version of the host $\bS$. To simplify the exposition, the degradation
arises via a deterministic symbolwise mapping $h~:~\calS \to \calS^d$.
The sequence $s^d = h(s)$ could represent a coarse version of $\bs$,
or some other features of $\bs$. Two special cases are {\em private
fingerprinting} where $S^d = S$, and {\em public fingerprinting}
where $S^d = \emptyset$. 
The decoder produces an estimate
\begin{equation}
   \hat{\calK} = g_N(\bY,\bS^d,V)
\label{eq:detect all}
\end{equation}
of the coalition. A possible decision is the empty set, $\hat{\calK} = \emptyset$,
which is the reasonable choice when an accusation would be unreliable.
To summarize, we have

\begin{definition}
A {\em randomized} rate-$R$ length-$N$ fingerprinting code $(f_N, g_N)$ with
embedding distortion $D_1$ is a pair of encoder mapping 
$f_N~:~\calS^N \times \calV_N \times \{1, 2, \cdots, 2^{NR} \} \to \calX^N$
and decoder mapping
$g_N~:~\calY^N \times (\calS^d)^N \times \calV_N \to \{1, 2, \cdots, 2^{NR} \}^\star$.
\label{def:deterministic-code}
\end{definition}

The randomization is via the secret key $V$ and can take the form of permutations
of the symbol positions $\{1,2,\cdots,N\}$, permutations of the $2^{NR}$ fingerprint assignments,
and an auxiliary time-sharing sequence, as in \cite{Moulin03}---\cite{Moulin08}, \cite{Tardos03}.

We now state the attack models and define the error probabilities,
capacities, and error exponents.

\subsection{Collusion Channels}

The conditional type $p_{\by|\bx_{\calK}}$ is a random variable whose conditional distribution
given $\bx_{\calK}$ depends on the collusion channel $p_{\bY|\bX_{\calK}}$. 
Our fidelity constraint on the coalition is of the general form
\begin{equation}
  Pr[p_{\by|\bx_{\calK}} \in \scrW_K] = 1, 
\label{eq:WK}
\end{equation}
where $\scrW_K$ is a convex subset of $\scrP_{Y|X_{\calK}}$.
That is, the empirical conditional p.m.f. of the pirated copy given the marked copies
is restricted. Examples of $\scrW_K$ are given in \cite{Moulin08}, including
hard distortion constraints on the coalition:
\begin{equation}
   \scrW_K = \left\{ p_{Y|X_{\calK}} ~:~ \sum_{x_{\calK},y} \,p_{X_{\calK}}(x_{\calK})
	\,p_{Y|X_{\calK}}(y|x_{\calK}) \,\eE_\phi \,d_2(\phi(x_{\calK}),y) \le D_2 \right\}
\label{eq:expected-D2}
\end{equation}
where $\phi~:\calX^K \to \calS$ is a (possible randomized) permutation-invariant estimator
$\hat{S}=\phi(X_{\calK})$ of each host signal sample based on the corresponding marked samples;
$d_2 ~:~ \calS \to \calY$ is the coalition's distortion function;
$p_{X_{\calK}}$ is a reference p.m.f.; and $D_2$ is the maximum allowed distortion.
Another possible choice for $\scrW_K$ is obtained using the Boneh-Shaw constraint
\cite{Boneh95,Moulin08}.

{\bf Fair Coalitions.}
Denote by $\pi$ a permutation of the elements of $\calK$.
The set of fair, feasible collusion channels is the subset of $\scrW_K$
consisting of permutation-invariant channels:
\begin{equation}
   \scrW_K^{fair} = \left\{ p_{Y|X_{\calK}} \in \scrW_K
	~:~ p_{Y|X_{\pi(\calK)}} = p_{Y|X_{\calK}} , \;\forall \pi \right\} .
\label{eq:WK-fair}
\end{equation}
The collusion channel $p_{\bY|\bX_{\calK}}$ is said to be fair if
$Pr[p_{\by|\bx_{\calK}} \in \scrW_K^{fair}]=1$.
For any fair collusion channel, the conditional type $p_{\by|\bx_{\calK}}$
is invariant to permutations of the colluders.

{\bf Strongly exchangeable collusion channels} \cite{Somekh05}.
Now denote by $\pi$ a permutation of the samples of a length-$N$ sequence.
For strongly exchangeable channels, $p_{\bY|\bX_{\calK}}(\pi\by|\pi\bx_{\calK})$
is independent of $\pi$, for every $(\bx_{\calK},\by)$. The channel is defined by
a probability assignment $Pr[T_{\by|\bx_{\calK}}]$ on the conditional type classes.
The distribution of $\bY$ conditioned on $\bY \in T_{\by|\bx_{\calK}}$ is uniform:
\begin{equation}
   p_{\bY|\bX_{\calK}}(\tby|\bx_{\calK})
	= \frac{Pr[T_{\by|\bx_{\calK}}]}{|T_{\by|\bx_{\calK}}|} ,
		\quad \forall \tby \in T_{\by|\bx_{\calK}} .
\label{eq:exch}
\end{equation}

\subsection{Error Probabilities}
\label{sec:probabilityOfError}

Let $\calK$ be the actual coalition and $\hat{\calK} = g_N(\bY,\bS^d,V)$ the decoder's
output. The three error probabilities of interest in this paper are
the probability of false positives (one or more innocent users are accused),
\[ P_{FP}(f_N,g_N,p_{\bY|\bX_{\calK}}) = Pr[\hat{\calK} \setminus \calK \ne \emptyset] , \]
the probability of failing to catch a single colluder,
\[ P_e^{one}(f_N,g_N,p_{\bY|\bX_{\calK}}) = Pr[\hat{\calK} \cap \calK = \emptyset] , \]
and the probability of failing to catch the full coalition: 
\[ P_e^{all}(f_N,g_N,p_{\bY|\bX_{\calK}}) = Pr[\calK \not\subseteq \hat{\calK}] . \]
These three probabilities are obtained by averaging over $\bS$, $V$, and the output of
the collusion channel $p_{\bY|\bX_{\calK}}$.
In each case the worst-case probability is denoted by
\begin{equation}
   P_e(f_N,g_N,\scrW_K) = \max_{p_{\bY|\bX_{\calK}}} \,P_e(f_N,g_N,p_{\bY|\bX_{\calK}})
\label{eq:Pe-worst}
\end{equation}
where $P_e$ denotes either $P_{FP}$, $P_e^{one}$ or $P_e^{all}$,
and the maximum is over all feasible collusion channels, i.e., such that
(\ref{eq:WK}) holds.

\subsection{Capacity and Random-Coding Exponents}

\begin{definition}
A rate $R$ is achievable for embedding distortion $D_1$, collusion class $\scrW_K$,
and {\bf detect-one} criterion if there exists a sequence of $(N, \lceil 2^{NR} \rceil )$
randomized codes $(f_N,g_N)$ with maximum embedding distortion $D_1$, such that both
$P_{e,N}^{one}(f_N,g_N,\scrW_K)$ and $P_{FP,N}(f_N,g_N,\scrW_K)$ vanish as $N \to \infty$.
\label{def:detect-one}
\end{definition}

\begin{definition}
A rate $R$ is achievable for embedding distortion $D_1$, collusion class $\scrW_K$,
and {\bf detect-all} criterion if there exists a sequence of $(N, \lceil 2^{NR} \rceil )$
randomized codes $(f_N,g_N)$ with maximum embedding distortion $D_1$, such that both
$P_{e,N}^{all}(f_N,g_N,\scrW_K)$ and $P_{FP,N}(f_N,g_N,\scrW_K)$ vanish as $N \to \infty$.
\label{def:detect-all}
\end{definition}

\begin{definition}
Fingerprinting capacities $C^{one}(D_1,\scrW_K)$ and $C^{all}(D_1,\scrW_K)$ 
are the suprema of all achievable rates with respect to the detect-one and
detect-all criteria, respectively.
\label{def:C}
\end{definition}

For random codes the error exponents corresponding to (\ref{eq:Pe-worst}) are defined as
\begin{equation}
   E^{\{one,all,FP\}}(R,D_1,\scrW_K) = \liminf_{N \to \infty} \left[ - \frac{1}{N} 
	\log P_e^{\{one,all,FP\}}(f_N, g_N, \scrW_K) \right] .
\label{eq:exp}
\end{equation}
We have $C^{all}(D_1,\scrW_K) \le C^{one}(D_1,\scrW_K)$
and $E^{all}(R,D_1,\scrW_K) \le E^{one}(R,D_1,\scrW_K)$
because an error event for the detect-one problem is also an error event
for the detect-all problem.

\section{Overview of Random-Coding Scheme}
\label{sec:encoder}
\setcounter{equation}{0}

A brief overview of our scheme is given in this section.
The decoders will be specified later.
The scheme is designed to achieve a false-positive error exponent equal to $\Delta$
and assumes a nominal value $K_{nom}$ for coalition size.
Two arbitrarily large integers $L_w$ and $L_u$ are selected, defining alphabets
$\calW = \{1,2,\cdots,L_w\}$ and $\calU = \{1,2,\cdots,L_u\}$, respectively.
The parameters $\Delta, K_{nom}, L_w, L_u$ are used to identify a certain optimal
type class $T_{\bw}^*$ and conditional type classes $T_{U|S^d W}^*(\bs^d,\bw)$,
$T_{U|SW}^*(\bs,\bw)$ and $T_{X|USW}^*(\bu,\bs,\bw)$ for every possible $(\bu,\bs,\bw)$.
Optimality is defined relative to either the thresholding decoder of
Sec.~\ref{sec:simple} or the joint decoder of Sec.~\ref{sec:joint}.
The secret key $V$ consists of a random sequence $\bW \in T_{\bw^*}$ and the
collection (\ref{eq:code}) of random codebooks indexed by $\bs^d,\bw,\lambda$.

\subsection{Codebook}

A random constant-composition code
\begin{equation}
   \calC(\bs^d,\bw,\lambda) = \{\bu(l,m,\lambda),
	\;\,1 \le l \le 2^{N\rho(\lambda)} ,\;1 \le m \le 2^{NR}\}
\label{eq:code}
\end{equation}
is generated for each pair of sequences $(\bs^d,\bw) \in (\calS^d)^N \times T_{\bw}^*$
and conditional type $\lambda \in \scrP_{S|S^d W}^{[N]}$ by drawing $2^{N[R+\rho(\lambda)]}$
random sequences independently and uniformly from an optimized conditional type class
$T_{U|S^d W}^*(\bs^d,\bw)$, and arranging them into an array with $2^{NR}$ columns and
$2^{N\rho(\lambda)}$ rows. Similarly to \cite{Moulin07} (see Fig.~2 therein), we refer to
$\rho(\lambda)$ as the depth parameter of the array.

\subsection{Encoding Scheme}

Prior to encoding, a sequence $\bW \in \calW^N$ is drawn independently of $\bS$
and uniformly from $T_{\bw}^*$, and shared with the receiver.
Given $(\bS,\bW)$, the encoder determines the conditional type $\lambda = p_{\bs|\bs^d\bw}$
and performs the following  two steps for each user $1 \le m \le 2^{NR}$.
\begin{enumerate}
\item Find $l$ such that $\bu(l,m,\lambda) \in \calC(\bs^d,\bw,\lambda) \bigcap
	T_{U|SW}^*(\bs,\bw)$. If more than one such $l$ exists, pick one of them randomly
	(with uniform distribution). Let $\bu = \bu(l,m,\lambda)$.
	If no such $l$ can be found, generate $\bu$ uniformly
	from the conditional type class $T_{U|SW}^*(\bs,\bw)$.
\item Generate $\bX_m$ uniformly distributed over the conditional type class 
	$T_{X|USW}^*(\bu,\bs,\bw)$, and assign this marked sequence to user $m$.
\end{enumerate}

\subsection{Worst Collusion Channel}

The fingerprinting codes used in this paper are randomly-modulated (RM) codes
\cite[Def.~2.2]{Moulin08}. For such codes we have the following proposition, 
which is a straightforward variation of \cite[Prop.~2.1]{Moulin08} with $\bS^d$
in place of $\bS$ at the decoder.
\begin{proposition}
For any RM code $(f_N,g_N)$, the maximum of the error probability criteria
(\ref{eq:Pe-worst}) over all feasible $p_{\bY|\bX_{\calK}}$ is achieved by a
strongly exchangeable collusion channel, as defined in (\ref{eq:exch}).
\label{prop:exch}
\end{proposition}
To derive error exponents for such channels, it suffices to use
the following upper bound:
\begin{equation}
   p_{\bY|\bX_{\calK}}(\tby|\bx_{\calK})
	= \frac{Pr[T_{\by|\bx_{\calK}}]}{|T_{\by|\bx_{\calK}}|} 
	\le \frac{1}{|T_{\by|\bx_{\calK}}|} \,\mathds1_{\{ p_{\by|\bx_{\calK}} \in \scrW_K \} } ,
		\quad \forall \,\tby \in T_{\by|\bx_{\calK}}
\label{eq:bound2}
\end{equation}
which holds uniformly over all feasible probability assignments 
to conditional type classes $T_{\by|\bx_{\calK}}$.

\subsection{Encoding and Decoding Errors}

The array depth parameter $\rho(\lambda)$ takes the form
\[ \rho(\lambda) = I(\bu;\bs|\bs^d,\bw) + \epsilon \]
where $\bu$ is any element of $T_{U|SW}^*(\bs,\bw)$, and $\epsilon > 0$
is an arbitrarily small number.
The analysis shows that given any $(\bs,\bw)$, the probability of encoding errors
vanishes doubly exponentially.

The analysis also shows that the decoding error probability is dominated by a single
joint type class $T_{\by\bu\bs\bw}$. Denote by $(\by,\bu,\bs,\bw)$ an arbitrary representative
of that class. The normalized logarithm of the size of the array is given by
\[ R + \rho(\lambda) = I(\bu;\by|\bs^d,\bw) - \Delta , \]
and the probability of false positives vanishes as $2^{-N\Delta}$.

\section{Threshold Decoder}
\label{sec:simple}
\setcounter{equation}{0}

\subsection{Decoding}

The decoder has access to $(\by,\bs^d,\bw)$ but does not know the
conditional type $\lambda = p_{\bs|\bs^d\bw}$ realized at the encoder.
The decoder evaluates the users one at a time and makes an innocent/guilty decision
on each user {\em independently of the other users}. Specifically, the receiver outputs
an estimated coalition $\hat{\calK}$ if and only if $\hat{\calK}$ satisfies
the following condition:
\begin{equation}
  \forall m \in \hat{\calK} ~:\quad \max_{\lambda \in \scrP_{S|S^d W}^{[N]}} 
  \max_{1 \le l \le 2^{N\rho(\lambda)} } I(\bu(l,m,\lambda);\by|\bs^d\bw)
			- \rho(\lambda) > R+\Delta .
\label{eq:decision-simple}
\end{equation}
If no such $\hat{\calK}$ is found, the receiver outputs $\hat{\calK} = \emptyset$.
This decoder outputs all user indices whose empirical mutual information
score, penalized by $\rho(\lambda)$, exceeds the threshold $R+\Delta$.

Observe that the maximizing $\lambda$ in (\ref{eq:decision-simple}) {\em may} depend
on $m$. With high probability, this event implies a decoding error. Improvements
can only be obtained using a more complex joint decoder, as in Sec.~\ref{sec:joint}.

\subsection{Error Exponents}

Define the following set of conditional p.m.f.'s for $(XU)_{\calK} \triangleq (X_{\calK}, U_{\calK})$
given $(S,W)$:
\[ \calM(p_{XU|SW}) = \{ p_{(XU)_{\calK}|SW} ~:~ p_{X_m U_m|SW} = p_{XU|SW} , \,m \in \calK \} , \]
i.e., the conditional marginal p.m.f. $p_{XU|SW}$ is the same for each $(X_m,U_m), \forall m \in \calK$.
Also define the sets
\begin{eqnarray}
   \scrP_{XU|SW}(p_{SW},L_w,L_u,D_1) 
	& = & \left\{ p_{XU|SW} ~:~ \eE[d(S,X)] \le D_1 \right\} , \nonumber \\
   \scrP_{(XU)_{\calK} W|S}(p_S,L_w,L_u,D_1) 
	& = & \left\{ p_{(XU)_{\calK} W|S} = p_W \,\prod_{k\in\calK} p_{X_k U_k|SW} \right. \nonumber \\
	& & \left. :~ p_{X_1 U_1|SW} = \cdots = p_{X_K U_K|SW} ,
	\;\;\mathrm{and}\;\; \eE[d(S,X_1)] \le D_1 \right\}
\label{eq:PXUS-set}
\end{eqnarray}
where in (\ref{eq:PXUS-set}) the random variables $(X_k,U_k),\,k\in\calK$,
are conditionally i.i.d. given $(S,W)$.

Define for each $m\in\calK$ the set of conditional p.m.f.'s
\begin{eqnarray}
   \lefteqn{\scrP_{Y(XU)_{\calK}|SW}(p_W, \tp_{S|W}, p_{XU|SW}, \scrW_K, R, L_w,L_u, m)} \nonumber\\
	& \triangleq & \left\{ \tp_{Y(XU)_{\calK}|SW}\,: ~\tp_{(XU)_{\calK}|SW} \in \calM(p_{XU|SW}), 
		~\tp_{Y|X_{\calK}} \in \scrW_K , \right. \nonumber\\
	& & \qquad \left. I_{p_W \tp_{S|W} \tp_{Y(XU)_{\calK}|SW}}(U_m;Y|S^d W)
		- I_{p_W \tp_{S|W} p_{XU|SW}}(U;S|S^d W) \le R \right\}
\label{eq:set-m-simple}
\end{eqnarray}
and the {\em pseudo sphere packing exponent}
\begin{eqnarray}
	\tilde{E}_{psp,m}(R,p_W,\tp_{S|W},p_{XU|SW},\scrW_K)
	& = & \min_{\tp_{Y(XU)_{\calK}|SW} \in \scrP_{Y(XU)_{\calK}|SW}
			(p_W, \tp_{S|W}, p_{XU|SW}, \scrW_K, R, L_w,L_u, m)} \nonumber \\
	& & \quad D(\tp_{Y(XU)_{\calK}|SW} \,\tp_{S|W} \| \tp_{Y|X_{\calK}} \,p_{XU|SW}^{\calK} \,p_S | p_W) .
\label{eq:Epsp-m-simple}
\end{eqnarray}
Taking the maximum and minimum of $\tilde{E}_{psp,m}$ above over $m\in\calK$,
we respectively define
\begin{eqnarray}
   \!\!\!\overline{\tilde{E}}_{psp}(R,L_w,L_u,p_W,\tp_{S|W},p_{XU|SW},\scrW_K)
		& = & \max_{m\in\calK} \tilde{E}_{psp,m}(R,L_w,L_u,p_W,\tp_{S|W},p_{XU|SW},\scrW_K) ,
																	\label{eq:Emax-tilde-simple} \\
   \!\!\!\underline{\tilde{E}}_{psp}(R,L_w,L_u,p_W,\tp_{S|W},p_{XU|SW},\scrW_K)
		& = & \min_{m\in\calK} \tilde{E}_{psp,m}(R,L_w,L_u,p_W,\tp_{S|W},p_{XU|SW},\scrW_K) .
																	\label{eq:Emin-tilde-simple}
\end{eqnarray}
For a fair coalition ($\scrW_K = \scrW_K^{fair}$), $\tilde{E}_{psp,m}$ is independent
of $m \in \calK$, and the two expressions above coincide. Define
\begin{eqnarray}
   E_{psp}(R,L_w,L_u,D_1,\scrW_K)
	& = & \max_{p_W \in \scrP_W} \min_{\tp_{S|W} \in \scrP_{S|W}}
	\max_{p_{XU|SW} \in \scrP_{XU|SW}(p_W \,\tp_{S|W},L_w,L_u,D_1)} \nonumber \\
	& & \quad \tilde{E}_{psp,1}(R,L_w,L_u,p_W,\tp_{S|W},p_{XU|SW},\scrW_K^{fair}) .
\label{eq:Epsp-simple}
\end{eqnarray}
Denote by $p_W^*$ and $p_{XU|SW}^*$ the maximizers in (\ref{eq:Epsp-simple}),
the latter to be viewed as a function of $\tp_{S|W}$. Both  $p_W^*$ and $p_{XU|SW}^*$
implicitly depend on $R$ and $\scrW_K^{fair}$.
Finally, define
\begin{eqnarray}
   \overline{E}_{psp}(R,L_w,L_u,D_1,\scrW_K) & = & \min_{\tp_{S|W} \in \scrP_{S|W}}
	\overline{\tilde{E}}_{psp}(R,L_w,L_u,p_W^*,\tp_{S|W},p_{XU|SW}^*,\scrW_K)
													\label{eq:Emax-simple} \\
   \underline{E}_{psp}(R,L_w,L_u,D_1,\scrW_K) & = & \min_{\tp_{S|W} \in \scrP_{S|W}}
	\underline{\tilde{E}}_{psp}(R,L_w,L_u,p_W^*,\tp_{S|W},p_{XU|SW}^*,\scrW_K) .
													\label{eq:Emin-simple}
\end{eqnarray}

The terminology {\em pseudo sphere-packing exponent} is used because despite its superficial
similarity to a real sphere-packing exponent, (\ref{eq:Epsp-m-simple}) does not provide
a fundamental asymptotic lower bound on error probability. 

\begin{theorem}
The decision rule (\ref{eq:decision-simple}) yields the following error exponents.
\begin{description}
\item[(i)] The false-positive error exponent is
	\begin{equation}
	   E_{FP}(R,D_1,\scrW_K,\Delta) = \Delta .
	\label{eq:E-FP-simple}
	\end{equation}
\item[(ii)] The detect-all error exponent is
	\begin{equation}
	   E^{all}(R,L_w,L_u,D_1,\scrW_K,\Delta)
		= \underline{E}_{psp}(R+\Delta,L_w,L_u,D_1,\scrW_K) .
	\label{eq:E-all-simple}
	\end{equation}
\item[(iii)] The detect-one error exponent is
	\begin{equation}
	   E^{one}(R,L_w,L_u,D_1,\scrW_K,\Delta)
		= \overline{E}_{psp}(R+\Delta,L_w,L_u,D_1,\scrW_K) .
	\label{eq:E-one-simple}
	\end{equation}
\item[(iv)] A fair collusion strategy is optimal under the detect-one error criterion:
	\[ E^{one}(R,L_w,L_u,D_1,\scrW_K,\Delta) = E^{one}(R,L_w,L_u,D_1,\scrW_K^{fair},\Delta) . \]
\item[(v)] The detect-one and detect-all error exponents 
	are the same when the colluders emply a fair strategy:
	$E^{one}(R,L_w,L_u,D_1,\scrW_K^{fair},\Delta) = E^{all}(R,L_w,L_u,D_1,\scrW_K^{fair},\Delta)$.
\item[(vi)] For $K = K_{nom}$, the supremum of all rates for which the detect-one
	error exponent of (\ref{eq:E-one-simple}) is positive is
	\begin{eqnarray}
		C^{thr}(D_1,\scrW_K) & = & C^{thr}(D_1,\scrW_K^{fair}) \nonumber \\
		& = & \lim_{L_w,L_u \to \infty} \;\max_{p_W \in \scrP_W}
			\;\max_{p_{XU|SW}  \in \scrP_{XU|SW}(p_W \,p_S, L_w,L_u,D_1)}
			\;\min_{p_{Y|X_{\calK}}  \in \scrW_K^{fair}} \nonumber \\
		& & \qquad [I(U;Y|S^d,W) - I(U;S|S^d,W)] .
	\label{eq:C1}
	\end{eqnarray}
\end{description}
\label{thm:simple}
\end{theorem}

\section{Joint Fingerprint Decoder}
\label{sec:joint}
\setcounter{equation}{0}

The fundamental improvement over the simple thresholding strategy for decoding in
Sec.~\ref{sec:simple} resides in the use of a joint decoding rule. Specifically,
the decoder maximizes a penalized empirical mutual information score over all
possible coalitions of any size.
The penalty depends on the conditional host sequence  type $p_{\bs|\bs^d\bw}$,
as in Sec.~\ref{sec:simple}, and is proportional to the size of the coalition,
as in \cite[Sec.~V]{Moulin08}.
We call this blind fingerprint decoder the {\em maximum doubly-penalized
mutual information} (M2PMI) decoder.

{\bf Mutual Information of $k$ Random Variables.}
The mutual information of $k$ random variables $X_1, \cdots, X_k$
is defined as the sum of their individual entropies minus
their joint entropy \cite[p.~57]{Csiszar81}
or equivalently, the divergence between their joint distribution
and the product of their marginals:
\begin{eqnarray}
   \oI(X_1;\cdots;X_k) & = & H(X_1) + \cdots + H(X_k) - H(X_1,\cdots,X_k)
												\label{eq:MI-k} \\
					 & = & D(p_{X_1 \cdots X_k} \| p_{X_1} \cdots p_{X_k}) . \nonumber
\end{eqnarray}
The symbol $\oI$ is used to distinguish it from ordinary mutual information $I$
between two random variables. Similarly one can define a conditional mutual
information $\oI(X_1;\cdots;X_k|Z) = \sum_i H(X_i|Z) - H(X_1,\cdots,X_k|Z)$
conditioned on $Z$, and an empirical mutual information
$\oI(\bx_1;\cdots;\bx_k|\bz)$ between $k$ sequences $\bx_1, \cdots, \bx_k$, conditioned
on $\bz$, as the conditional mutual information with respect to the joint type of
$\bx_1, \cdots, \bx_k, \bz$.
Some properties of $\oI$ are given in \cite[Sec.~V.A]{Moulin08}.

Recall that $\bx_{\calA}$ denotes $\{\bx_m, \,m \in \calA\}$ and that the codewords
in (\ref{eq:code}) take the form $\bu(l,m,\lambda)$.
In the following, we shall use the compact notation
$(\bx\bu)_{\calA} \triangleq (\bx_{\calA}, \bu_{\calA})$, and
\[ \bu(l_{\calA}, m_{\calA}, \lambda)
	\triangleq \{ \bu(l_{m_1},m_1,\lambda), \cdots,
	\bu(l_{m_{|\calA|}},m_{|\calA|},\lambda) \}
	\quad \mathrm{for~} \calA = \{ m_1, \cdots, m_{|\calA|} \} .
\]

\subsection{M2PMI Criterion}
\label{sec:MPMI}

Given $\by,\bs^d,\bw$, the decoder seeks the coalition size $k$, the conditional host
sequence type $\lambda \in \scrP_{S|S^d W}^{[N]}$, and the codewords $\bu(l,m,\lambda)$
in $\calC(\bs^d,\bw,\lambda)$ that maximize the M2PMI criterion below. The column indices $m\in\calK$,
corresponding to the decoded words form the decoded coalition $\hat{\calK}$.
If the maximizing $k$ in (\ref{eq:M2PMI}) is zero, the receiver outputs $\hat{\calK} = \emptyset$.

The {\em Maximum Doubly-Penalized Mutual Information} criterion is defined as
\begin{equation}
   \max_{k \ge 0} M2PMI(k)
\label{eq:M2PMI}
\end{equation}
where
\begin{equation}
   M2PMI(k) = \left\{ \begin{array}{ll} 0 & :~\mathrm{if~} k = 0 \\
	\underset{\lambda \in \scrP_{S|S^d W}^{[N]}}{\max} \;\underset{\bu_{\calK}\in\calC^k(\bs^d,\bw,\lambda)}{\max}
	\left[\oI(\bu_{\calK};\by|\bs^d\bw) - k (\rho(\lambda)+R+\Delta) \right]
				& :~\mathrm{if~} k=1,2,\cdots
				\end{array} \right.
\label{eq:M2PMI-k}
\end{equation}


\subsection{Properties}
\label{sec:joint-properties}

The following lemma shows that
1) each subset of the estimated coalition is significant, and 
2) any further extension of the coalition would fail a significance test. 
The proof parallels that of Lemma~5.1 in \cite{Moulin08} and is therefore omitted.
\begin{lemma}
Let $\hat{\calK}$, $\lambda$, $l_{\hat{\calK}}$ achieve the maximum in (\ref{eq:M2PMI-k})
(\ref{eq:M2PMI}), i.e., $\bu_{\hat{\calK}} = \bu(l_{\hat{\calK}},m_{\hat{\calK}},\lambda)$.
Then for each subset of the estimated coalition $\hat{\calK}$, we have
\begin{equation}
  \forall \calA \subseteq \hat{\calK} ~:\quad
  \oI(\bu(l_{\calA}, m_{\calA}, \lambda);
		\by\bu(l_{\hat{\calK}\setminus\calA}, m_{\hat{\calK}\setminus\calA}, \lambda)\,|\bs^d\bw)
		> |\calA| \,(\rho(\lambda) + R+\Delta) .
\label{eq:property1}
\end{equation}
Moreover, for every $\calA$ disjoint with $\hat{\calK}$,
\begin{equation}
  \oI(\bu(l_{\calA}, m_{\calA}, \lambda);
	\by\bu(l_{\hat{\calK}}, m_{\hat{\calK}}, \lambda)\,|\bs^d\bw)
		\le |\calA| \,(\rho(\lambda) + R+\Delta) .
\label{eq:property2}
\end{equation}
\label{lem:optimalK}
\end{lemma}

\subsection{Error Exponents}
\label{sec:joint-exp}

Define for each $\calA \subseteq \calK$ the set of conditional p.m.f.'s
\begin{eqnarray}
   \lefteqn{\scrP_{Y(XU)_{\calK}|SW}(p_W,\tp_{S|W}, p_{XU|SW}, \scrW_K, R, L_w,L_u, \calA)} \nonumber \\
	& \triangleq & \left\{ \tp_{Y(XU)_{\calK}|SW}\,: ~\tp_{(XU)_{\calK}|SW} \in \calM(p_{XU|SW}) ,
																		\right. \nonumber \\
	& & \quad \left. \frac{1}{|\calA|} \oI_{p_W\,\tp_{S|W} \,\tp_{Y(XU)_{\calK}|SW}}
			(U_{\calA};YU_{\calK\setminus\calA}|S^d,W) \le I_{p_W\,\tp_{S|W} \,p_{XU|SW}}(U;S|S^d,W)+R \right\}
\label{eq:set-A}
\end{eqnarray}
and the {\em pseudo sphere packing exponent} 
\begin{eqnarray}
   \tilde{E}_{psp,\calA}(R,L_w,L_u,p_W,\tp_{S|W},p_{XU|SW},\scrW_K)
	& = & \min_{\tp_{Y(XU)_{\calK}|SW} \in \scrP_{Y(XU)_{\calK}|SW}
		(p_W, \tp_{S|W}, p_{XU|SW}, \scrW_K, R, L_w,L_u, \calA)} \nonumber \\
	& & D(\tp_{Y(XU)_{\calK}|SW} \,\tp_{S|W} \| \tp_{Y|X_{\calK}} \,\tp_{(XU)_{\calK}|SW} \,p_S \,|p_W) .
\label{eq:Epsp-A}
\end{eqnarray}
Taking the maximum \footnote{
   The property that $\calK$ achieves $\max_{\calA \subseteq \calK} \tilde{E}_{psp,\calA}$
   is established in the proof of Theorem~\ref{thm:joint}, Part (iv).
}
and the minimum of $\tilde{E}_{psp,\calA}$ above over all subsets $\calA$ of $\calK$,
we define
\begin{eqnarray}
   \!\!\!\!\!\overline{\tilde{E}}_{psp}(R,L_w,L_u,p_W,\tp_{S|W},p_{XU|SW},\scrW_K)
		& = & \tilde{E}_{psp,\calK}(R,L_w,L_u,p_W,\tp_{S|W},p_{XU|SW},\scrW_K) ,
																	\label{eq:Emax-tilde} \\
   \!\!\!\!\!\underline{\tilde{E}}_{psp}(R,L_w,L_u,p_W,\tp_{S|W},p_{XU|SW},\scrW_K)
		& = & \min_{\calA \subseteq \calK}
				\tilde{E}_{psp,\calA}(R,L_w,L_u,p_W,\tp_{S|W},p_{XU|SW},\scrW_K) .
																	\label{eq:Emin-tilde}
\end{eqnarray}
Now define
\begin{eqnarray}
   E_{psp}(R,L_w,L_u,D_1,\scrW_K) & = & \max_{p_W \in \scrP_W} \min_{\tp_{S|W} \in \scrP_{S|W}}
	\max_{p_{XU|SW} \in \scrP_{XU|SW}(p_W,\tp_{S|W},L_w,L_u,D_1)} \nonumber \\
	& & \quad \tilde{E}_{psp,\calK}(R,L_w,L_u,p_W,\tp_{S|W},p_{XU|SW},\scrW_{K_{nom}}^{fair}) .
\label{eq:Epsp}
\end{eqnarray}
Denote by $p_W^*$ and $p_{XU|SW}^*$ the maximizers in (\ref{eq:Epsp}),
where the latter is to be viewed as a function of $\tp_{S|W}$.
Both  $p_W^*$ and $p_{XU|SW}^*$ implicitly depend on $R$ and $\scrW_K^{fair}$.
Finally, define
\begin{eqnarray}
   \overline{E}_{psp}(R,L_w,L_u,D_1,\scrW_K) & = & \min_{\tp_{S|W} \in \scrP_{S|W}}
	\overline{\tilde{E}}_{psp}(R,L_w,L_u,p_W^*,\tp_{S|W},p_{XU|SW}^*,\scrW_K) ,
													\label{eq:Emax} \\
   \underline{E}_{psp}(R,L_w,L_u,D_1,\scrW_K) & = & \min_{\tp_{S|W} \in \scrP_{S|W}}
	\underline{\tilde{E}}_{psp}(R,L_w,L_u,p_W^*,\tp_{S|W},p_{XU|SW}^*,\scrW_K) .
													\label{eq:Emin}
\end{eqnarray}

\begin{theorem}
The decision rule (\ref{eq:M2PMI}) yields the following error exponents.
\begin{description}
\item[(i)] The false-positive error exponent is
	\begin{equation}
	   E_{FP}(R,D_1,\scrW_K,\Delta) = \Delta .
	\label{eq:E-FP}
	\end{equation}
\item[(ii)] The detect-all error exponent is
	\begin{equation}
	   E^{all}(R,L_w,L_u,D_1,\scrW_K,\Delta) = \underline{E}_{psp}(R+\Delta,L_w,L_u,D_1,\scrW_K) .
	\label{eq:E-all}
	\end{equation}
\item[(iii)] The detect-one error exponent is
	\begin{equation}
	   E^{one}(R,L_w,L_u,D_1,\scrW_K,\Delta) = \overline{E}_{psp}(R+\Delta,L_w,L_u,D_1,\scrW_K) .
	\label{eq:E-one}
	\end{equation}
\item[(iv)] $E^{one}(R,L_w,L_u,D_1,\scrW_K,\Delta) = E^{one}(R,L_w,L_u,D_1,\scrW_K^{fair},\Delta)$.
\item[(v)]  $E^{all}(R,L_w,L_u,D_1,\scrW_K^{fair},\Delta) = E^{one}(R,L_w,L_u,D_1,\scrW_K^{fair},\Delta)$.
\item[(vi)] If $K = K_{nom}$, the supremum of all rates for which the error exponent of
	(\ref{eq:E-one}) and (\ref{eq:E-all}) are positive is
	\begin{eqnarray}
	\underline{C}^{one}(D_1,\scrW_K)
	& = & \underline{C}^{one}(D_1,\scrW_K^{fair}) \nonumber \\
	& = & \lim_{L_w,L_u \to \infty} \;\max_{p_W \in \scrP_W} 
			\;\max_{p_{(XU)_{\calK}|SW} \in \scrP_{(XU)_{\calK}|SW}(p_W,p_S,L_w,L_u,D_1)}
			\;\min_{p_{Y|X_{\calK}} \in \scrW_K^{fair}} \nonumber \\
	& & \qquad \left[\frac{1}{K} I(U_{\calK};Y|S^d,W)-I(U;S|S^d,W)\right]
	\label{eq:C-achieve-one}
	\end{eqnarray}
	under the ``detect-one'' criterion, and by
	\begin{eqnarray}
	\underline{C}^{all}(D_1,\scrW_K)
    & = & \lim_{L_w,L_u \to \infty} \;\max_{p_W \in \scrP_W}
			\;\max_{p_{(XU)_{\calK}|SW} \in \scrP_{(XU)_{\calK}|SW}(p_W,p_S,L_w,L_u,D_1)}
			\;\min_{p_{Y|X_{\calK}}  \in \scrW_K} 		\nonumber \\
	& & \qquad \left[ \min_{\calA \subseteq \calK}
		\;\frac{1}{|\calA|} I(U_{\calA} ; Y \,|S^d, W, U_{\calK\setminus\calA}) - I(U;S|S^d,W) \right]
	\label{eq:C-achieve-all}
	\end{eqnarray}
	under the ``detect-all'' criterion. If the colluders select a fair collusion channel,
	as is their collective interest, the minimization is restricted to
	$\scrW_K^{fair}$ in (\ref{eq:C-achieve-all}), and then
	\[ \underline{C}^{all}(D_1,\scrW_K) = \underline{C}^{one}(D_1,\scrW_K) . \]
\end{description}
\label{thm:joint}
\end{theorem}

For the special case of private fingerprinting ($S^d = S$),
the term $I(U;S|S^d,W)$ in (\ref{eq:C-achieve-one}) is zero. Since
$I(U_{\calK};Y|S,W) \le I((XU)_{\calK};Y|S,W)$, it suffices to choose
$L_u=|\calX|$ and $U=X$ to achieve the maximum in (\ref{eq:C-achieve-one}).
The resulting expression coincides with the capacity formula
in \cite[Theorem~3.2]{Moulin08}. Similarly to the single-user case \cite{Moulin07},
when $U=X$ the binning scheme is degenerate.

\subsection{Bounded Coalition Size}

Assume now that $K$ is known not exceed some maximum value $K_{\max}$.
The same random coding scheme can be used.
In the evaluation of the M2PMI criterion of (\ref{eq:M2PMI}),
the maximization is now limited to $0 \le k \le K_{\max}$.
In Lemma~\ref{lem:optimalK}, property (\ref{eq:property1}) holds, and
property (\ref{eq:property2}) now holds for every $\calA$ disjoint with $\hat{\calK}$,
and of size $|\calA| \le K_{\max} - |\hat{\calK}|$. Following the derivation
of the error exponents in the appendix, we see that these exponents remain 
the same as those given by Theorem~\ref{thm:joint}.

{\bf Blind watermarking}. The case $K_{\max}=1$ represents blind watermark decoding
with a guarantee that the false-positive exponent is at least equal to $\Delta$.
In this scenario, there is no need for a time-sharing
sequence $\bw$, and the decoder's input $\by$ is either
an unwatermarked sequence ($K=0$) or a watermarked sequence ($K=1$).
The M2PMI criterion of (\ref{eq:M2PMI-k}) reduces to
\[ M2PMI(k)=\max_\lambda \max_{\bu \in \calC(\bs^d)}
I(\bu;\by|\bs^d) - (\rho(\lambda)+R+\Delta) \quad \mathrm{for~} k=1. \]
The resulting false-positive and false-negative exponents
are given by $\Delta$ and $E_{psp}(R+\Delta,0,L_u,D_1,\scrW_K)$, respectively.

\section{Upper Bounds on Public Fingerprinting Capacity}
\label{sec:C-public}
\setcounter{equation}{0}

Deriving public fingerprinting capacity is a challenge because the capacity
region for the Gel'fand-Pinsker version of the MAC is still unknown, in fact
an outer bound for this region has yet to be established. Even in the case
of a MAC with side information {\em causally} available at the transmitter
but not at the receiver, the expressions for the inner and outer capacity regions
do not coincide \cite{Sigurjonsson05}. Likewise, the expression derived below
is an upper bound on public fingerprinting capacity under the detect-all criterion.

Recall the definition of the set $\scrP_{(XU)_{\calK} W|S}(p_S,L_w,L_u,D_1)$
in (\ref{eq:PXUS-set}), where $W$ and $U$ are random variables defined over alphabets
$\calW = \{1,2,\cdots,L_w\}$ and $\calU = \{1,2,\cdots,L_u\}$, respectively.
Here we define the larger set
\begin{eqnarray}
   \scrP_{(XU)_{\calK} W|S}^{outer}(p_S,L_w,L_u,D_1)
	& = & \left\{ p_{(XU)_{\calK} W|S} = p_W \,\left( \prod_{k\in\calK} p_{X_k|SW} \right)
		p_{U_{\calK}|X_{\calK} SW} : \right. \nonumber \\
	& & \left. \quad p_{X_1|SW} = \cdots = p_{X_K|SW} ,
	\;\;\mathrm{and}\;\; \eE[d(S,X_1)] \le D_1 \right\}
\label{eq:PXUS-outer-set}
\end{eqnarray}
where $X_k,\,k\in\calK$, are still conditionally i.i.d. given $(S,W)$
but the random variables $U_k,\,k\in\calK$, are generally conditionally dependent.

Define
\begin{eqnarray}
   \overline{C}_{L_w,L_u}^{all}(D_1,\scrW_K)
    & = & \max_{p_{(XU)_{\calK} W|S} \in \scrP_{(XU)_{\calK} W|S}^{outer}(p_S,L_w,L_u,D_1)}
		\;\min_{p_{Y|X_{\calK}} \in \scrW_K} \nonumber \\
	& & \qquad \qquad \min_{\calA \subseteq \calK}
		\;\frac{1}{|\calA|} \left[ I(U_{\calA};Y,S^d|U_{\calK\setminus\calA})
		- I(U_{\calA};S|U_{\calK\setminus\calA}) \right] .
\label{eq:CL-all}
\end{eqnarray}
Using the same derivation as in Lemma~2.1 of \cite{Moulin07}, it can be
shown that $\overline{C}_{L_w,L_u}^{all}(D_1,\scrW_K)$
is a nondecreasing function of $L_w$ and $L_u$ and converges to a finite limit.
Moreover, the gap to the limit may be bounded by a polynomial function of $L_w$
and $L_u$, see \cite[Sec.~3.5]{Moulin07} for a similar derivation.

\begin{theorem}
Public fingerprinting capacity is upper-bounded by
\begin{eqnarray}
   \overline{C}^{all}(D_1,\scrW_K) = \lim_{L_w, L_u \to \infty} \;\overline{C}_{L_w,L_u}^{all}(D_1,\scrW_K)
\label{eq:C-all}
\end{eqnarray}
under the ``detect-all'' criterion. 
\label{thm:public}
\end{theorem}
{\em Proof}: see appendix. 

We conjecture that the upper bound on capacity given by Theorem~\ref{thm:public}
is generally not tight. The insight here is that the upper bound remains
valid if the class of encoding functions is enlarged to include feedback
from the receiver: $X_{ki} = \tilde{f}_i(\bS,M_k,Y^{i-1})$ for $1 \le i \le N$.
It can indeed be  verified that all the inequalities in the proof
and the Markov chain properties hold. The question is now whether feedback can
increase public fingerprinting capacity. We conjecture the answer is yes,
because feedback is known to increase MAC capacity \cite{Gaarder75}.

We also make the stronger conjecture that the maximum over $p_{(XU)_{\calK}|SW}$ is
achieved by a p.m.f. that decouples the components $(X_k,U_k), \,k\in\calK$,
conditioned on $(S,W)$. 
If this is true, the set $\scrP_{(XU)_{\calK} W|S}^{outer}(p_S$, $L_w,L_u,D_1)$
in the formula (\ref{eq:CL-all}) can be replaced with
the smaller set $\scrP_{(XU)_{\calK} W|S}(p_S$, $L_w,L_u,D_1)$ of (\ref{eq:PXUS-set}),
and the random coding scheme of Sec.~\ref{sec:joint} is capacity-achieving.


\section{Conclusion}
\label{sec:conclusion}

We have proposed a communication model
and a random-coding scheme for blind fingerprinting. While a standard
binning scheme for communication with asymmetric side information
at the transmitter and the receiver may seem like a reasonable candidate,
such a scheme would be unable to trade false-positive error exponents
against false-negative error exponents.
Our proposed binning scheme combines two ideas.
The first is the use of a stacked binning scheme as in \cite{Moulin07},
which demonstrated the advantages (in terms of decoding error exponents)
of selecting codewords from an array whose size depends on the conditional
type of the host sequence. The second is the use of an auxiliary time-sharing
random variable as in \cite{Moulin08}.
The blind fingerprint decoders of Secs.~\ref{sec:simple} and \ref{sec:joint}
combine the advantages of both methods and provide positive error exponents
for a range of code rates.
The tradeoff between the two fundamental types of error probabilities
is determined by the value of the parameter $\Delta$.

\newpage

\appendices
\renewcommand{\theequation}{\Alph{section}.\arabic{equation}}

\section{Proof of Theorem~\ref{thm:simple}}
\label{Sec:SimpleProof}
\setcounter{equation}{0}

We derive the error exponents for the thresholding rule (\ref{eq:decision-simple}).
We have $\calW = \{ 1,2,\cdots,L_w\}$ and $\calU = \{ 1,2,\cdots,L_u\}$.
Fix some arbitrarily small $\epsilon > 0$. Define for all $m \in \calK$
\begin{eqnarray}
  \scrP_{Y(XU)_{\calK}|SW}^{[N]}(p_{\bw}, p_{\bs|\bw}, p_{\bx\bu|\bs\bw}, \scrW_K, R,L_w,L_u, m)
	& = & \left\{ p_{\by(\bx\bu)_{\calK}|\bs\bw}
		\,:~p_{(\bx\bu)_{\calK}|\bs\bw} \in \calM(p_{\bx\bu|\bs\bw}), \right. \nonumber \\
	& & \left. p_{\by|\bx_{\calK}} \in \scrW_K, ~I(\bu_m;\by|\bs^d \bw)
		\le \rho(p_{\bs|\bs^d\bw})+R \right\} \nonumber \\
  \breve{E}_{psp,m,N}(R,L_w,L_u,p_{\bw},p_{\bs|\bw}, p_{\bx\bu|\bs\bw},\scrW_K)
	& = & \min_{p_{\by(\bx\bu)_{\calK}|\bs\bw} \in \scrP_{Y(XU)_{\calK}|SW}^{[N]}
			(p_{\bw}, p_{\bs|\bw}, p_{\bx\bu|\bs\bw}, \scrW_K, R,L_w,L_u, m)} \nonumber \\
	&   & \quad D(p_{\by(\bx\bu)_{\calK}|\bs\bw}
			\| p_{\by|\bx_{\calK}} \,p_{\bx\bu|\bs\bw}^{\calK} | p_{\bs\bw} ) ,
														\label{eq:Epsp-m-simple-N} \\
   \hat{E}_{psp,m,N}(R,L_w,L_u,p_{\bw},p_{\bs|\bw}, p_{\bx\bu|\bs\bw},\scrW_K)
	& = & D(p_{\bs|\bw} \| p_S \,|\,p_{\bw}) + \breve{E}_{psp,m,N}(R,L_w,L_u, \nonumber\\
	& & \quad p_{\bw},p_{\bs|\bw}, p_{\bx\bu|\bs\bw},\scrW_K) \nonumber \\
	& = & \min_{p_{\by(\bx\bu)_{\calK}|\bs\bw} \in \scrP_{Y(XU)_{\calK}|SW}^{[N]}
			(p_{\bw}, p_{\bs|\bw}, p_{\bx\bu|\bs\bw}, \scrW_K, R,L_w,L_u, m)} \nonumber \\
	&   & \quad D(p_{\by(\bx\bu)_{\calK}|\bs\bw} \,p_{\bs|\bw}
			\| p_{\by|\bx_{\calK}} \,p_{\bx\bu|\bs\bw}^{\calK} \,p_S \,| p_{\bw} ) ,
															\label{eq:Ehat-m-simple-N} \\
   \overline{\hat{E}}_{psp,N}(R,L_w,L_u,p_{\bw},p_{\bs|\bw}, p_{\bx\bu|\bs\bw},\scrW_K)
	& = & \max_{m\in\calK} \hat{E}_{psp,m,N}(R,L_w,L_u,p_{\bw},p_{\bs|\bw}, p_{\bx\bu|\bs\bw},\scrW_K)
			\nonumber \\
			\label{eq:Emax-tilde-simple-N} \\
   \underline{\hat{E}}_{psp,N}(R,L_w,L_u,p_{\bw},p_{\bs|\bw}, p_{\bx\bu|\bs\bw},\scrW_K)
	& = & \min_{m\in\calK} \hat{E}_{psp,m,N}(R,L_w,L_u,p_{\bw},p_{\bs|\bw}, p_{\bx\bu|\bs\bw},\scrW_K)
			\nonumber \\
			\label{eq:Emin-tilde-simple-N}
\end{eqnarray}
where (\ref{eq:Ehat-m-simple-N}) is obtained by application of the chain rule for divergence.
Also define
\begin{eqnarray}
   E_{psp,N}(R,L_w,L_u,D_1,\scrW_K) & = & \max_{p_{\bw} \in \scrP_{W}^{[N]}}
	\min_{p_{\bs|\bw} \in \scrP_{S|W}^{[N]}}
	\max_{p_{\bx\bu|\bs\bw} \in \scrP_{XU|SW}^{[N]}(p_{\bw}\,p_{\bs|\bw},L_w,L_u,D_1)} \nonumber \\
	& & \hat{E}_{psp,1,N}(R,L_w,L_u,p_{\bw},p_{\bs|\bw}, p_{\bx\bu|\bs\bw},\scrW_{K_{nom}}^{fair}) .
\label{eq:Epsp-simple-N}
\end{eqnarray}
Denote by $p_{\bw}^*$ and $p_{\bx\bu|\bs\bw}^*$ the maximizers above, the latter viewed
as a function of $p_{\bs|\bw}$. Both maximizers depend implicitly on $R$ and $\scrW_{K_{nom}}^{fair}$.
Let
\begin{eqnarray}
   \overline{E}_{psp,N}(R,L_w,L_u,D_1,\scrW_K) & = & \min_{p_{\bs|\bw} \in \scrP_{S|W}^{[N]}}
	\overline{\hat{E}}_{psp,N}(R,L_w,L_u,p_{\bw}^*,p_{\bs|\bw}, p_{\bx\bu|\bs\bw}^*) ,
															\label{eq:Emax-simple-N} \\
   \underline{E}_{psp,N}(R,L_w,L_u,D_1,\scrW_K) & = & \min_{p_{\bs|\bw} \in \scrP_{S|W}^{[N]}}
	\underline{\hat{E}}_{psp,N}(R,L_w,L_u,p_{\bw}^*,p_{\bs|\bw}, p_{\bx\bu|\bs\bw}^*) .
															\label{eq:Emin-simple-N}
\end{eqnarray}
The exponents (\ref{eq:Ehat-m-simple-N})---(\ref{eq:Emin-simple-N})
differ from (\ref{eq:Epsp-m-simple})---(\ref{eq:Emin-simple}) in that
the optimizations are performed over conditional types instead of general conditional
p.m.f.'s. We have
\begin{eqnarray}
 	\lim_{N \to \infty} \overline{E}_{psp,N}(R,L_w,L_u,D_1,\scrW_K)
		& = & \overline{E}_{psp}(R,L_w,L_u,D_1,\scrW_K)
														\label{eq:lim-Emax-simple-N} \\
 	\lim_{N \to \infty} \underline{E}_{psp,N}(R,L_w,L_u,D_1,\scrW_K)
		& = & \underline{E}_{psp}(R,L_w,L_u,D_1,\scrW_K)
														\label{eq:lim-Emin-simple-N}
\end{eqnarray}
by continuity of the divergence and mutual-information functionals.

Consider the maximization over the conditional type $p_{\bx\bu|\bs\bw}$
in (\ref{eq:Epsp-simple-N}). 
As a result of this maximization, we may associate the following:
\begin{itemize}
\item to any $(\bs,\bw)$, a conditional type class 
	$T_{U|SW}^*(\bs,\bw) \triangleq T_{\bu|\bs\bw}^*$;
\item to any $(\bs^d,\bw)$, a conditional type class 
	$T_{U|S^d W}^*(\bs^d,\bw) \triangleq T_{\bu|\bs^d \bw}^*$;
\item to any $(\bs,\bw)$ and $\bu \in T_{U|SW}^*(\bs\bw)$, 
	a conditional type class $T_{X|USW}^*(\bu,\bs,\bw) \triangleq T_{\bx|\bu\bs\bw}^*$;
\item to any type $p_{\bs\bw}$, a conditional mutual information
	$I_{US|S^d W}^*(p_{\bs\bw}) \triangleq  I(\bu;\bs|\bs^d,\bw)$ where
	$\bu,\bs,\bw$ are any three sequences with joint type $p_{\bu|\bs\bw}^* p_{\bs\bw}$.
\end{itemize}

{\bf Codebook}.
Define the function
\[ \rho(p_{\bs|\bs^d \bw}) = I_{US|S^d W}^*(p_{\bs\bw}) + \epsilon ,
	\quad \forall p_{\bs\bw} \in \scrP_{SW}^{[N]} . \]
A random constant-composition code
\[ \calC(\bs^d,\bw,p_{\bs|\bs^d \bw}) = \{\bu(l,m,p_{\bs|\bs^d \bw}),
	\;\,1 \le l \le \exp_2 \{ N\rho(p_{\bs|\bs^d \bw}) ,\;1 \le m \le 2^{NR}\} \]
is generated for each $\bs^d \in (\calS^d)^N$, $\bw \in T_{\bw}^*$, and
$p_{\bs|\bs^d \bw} \in \scrP_{S|S^d W}^{[N]}$ by drawing
$\exp_2 \{ N(R+\rho(p_{\bs|\bs^d \bw})) \}$ random sequences independently and
uniformly from the conditional type class $T_{U|S^d W}^*(\bs^d,\bw)$, and arranging
them into an array with $2^{NR}$ columns and $\exp_2 \{ N\rho(p_{\bs|\bs^d \bw}) \}$
rows. 

{\bf Encoder}.
Prior to encoding, a sequence $\bW \in \calW^N$ is drawn independently of $\bS$
and uniformly from $T_{\bw}^*$, and shared with the receiver.
Given $(\bS,\bW)$, the encoder determines the conditional type $p_{\bs|\bs^d \bw}$
and performs the following two steps for each user $1 \le m \le 2^{NR}$.
\begin{enumerate}
\item Find $l$ such that $\bu(l,m,p_{\bs|\bs^d\bw}) \in \calC(\bS^d,\bW,p_{\bs|\bs^d\bw})
	\bigcap T_{U|SW}^*(\bs,\bw)$.
	If more than one such $l$ exists, pick one of them randomly (with uniform
	distribution). Let $\bu = \bu(l,m,p_{\bs|\bs^d\bw})$.
	If no such $l$ can be found, generate $\bu$ uniformly
	from the conditional type class $T_{U|SW}^*(\bs,\bw)$.
\item Generate $\bX_m$ uniformly distributed over the conditional type class 
	$T_{X|USW}^*(\bu,\bs,\bw)$.
\end{enumerate}

{\bf Collusion channel.} By Prop.~\ref{prop:exch}, it is sufficient to restrict
our attention to strongly exchangeable collusion channels in the error probability
analysis.

{\bf Decoder.} Given $(\by,\bs^d,\bw)$, the decoder outputs $\hat{\calK}$
if and only if (\ref{eq:decision-simple}) is satisfied.

{\bf Encoding errors.}
Analogously to \cite{Moulin07}, the probability of encoding errors vanishes
doubly exponentially with $N$ because $\rho(p_{\bs|\bs^d\bw}) > I(\bu;\bs|\bs^d\bw)$.
Indeed an encoding error for user $m$ arises under the following event:
\begin{eqnarray}
  \calE_m & = & \{ (\calC, \bs,\bw) ~:~ (\bu(l,m,p_{\bs|\bs^d\bw}) \in \calC
    ~\mathrm{and}~ \bu(l,m,p_{\bs|\bs^d\bw}) \notin T_{U|SW}^*(\bs,\bw))
    ~\mathrm{for}~ 1 \le l \le 2^{N\rho(p_{\bs|\bs^d\bw})} \} . \nonumber \\
\label{eq:E1}
\end{eqnarray}
The probability that a sequence $\bU$
uniformly distributed over $T_{U|S^d W}^*(\bs^d,\bw)$ also belongs
to $T_{U|SW}^*(\bs,\bw)$ is equal to $\exp_2 \{ -N I_{US|S^d W}^*(p_{\bs\bw}) \}$
on the exponential scale. Therefore the encoding error probability, conditioned
on type class $T_{\bs\bw}$, satisfies
\begin{eqnarray}
   Pr[\calE_m | (\bS,\bW) \in T_{\bs\bw}]
    & = & \left( 1 - \frac{|T_{U|SW}^*(\bS,\bW)|}{|T_{U|S^d W}^*(\bS^d,\bW)|} \right)
        ^{2^{N\rho(p_{\bs|\bs^d\bw})}} \nonumber \\
    & \doteq & ( 1 - 2^{-N I_{US|S^d W}^*(p_{\bs\bw})} )^{2^{N\rho(p_{\bs|\bs^d\bw})}} \nonumber \\
    & \le & \exp \{ - \exp_2 ( N [\rho(p_{\bs|\bs^d\bw}) - I_{US|S^d W}^*(p_{\bs\bw})] ) \} \nonumber \\
    & = & \exp \{ - 2^{N \epsilon} \}
\label{eq:double-exp}
\end{eqnarray}
where the inequality follows from $1-a \le e^{-a}$.

The derivation of the decoding error exponents is based on the following two
asymptotic equalities which are special cases of (\ref{eq:I-rc}) and (\ref{eq:I-sp})
established in Lemma~\ref{lem:3properties}.

\noindent
1) Fix $\by,\bs^d,\bw$ and draw $\bu$ uniformly from some fixed type class,
independently of $(\by,\bs^d,\bw)$. Then
\begin{equation}
   Pr[I(\bu;\by|\bs^d\bw) \ge \nu] \doteq 2^{-N \nu} .
\label{eq:I-rc-simple}
\end{equation}
2) Given $\bs,\bw$, draw $(\bx_k, \bu_k), \,k\in\calK$, i.i.d. uniformly from a conditional type
class $T_{\bx\bu|\bs\bw}$, and then draw $\by$ uniformly over a single conditional type class
$T_{\by|\bx_{\calK}}$.
For any $\nu > 0$, we have
\begin{equation}
   Pr[I(\bu_m;\by|\bs^d\bw) \le \rho(p_{\bs|\bs^d\bw}) + \nu] \doteq \exp_2
	\{ -N \breve{E}_{psp,m,N}(\nu,L_w,L_u,p_{\bw}^*, p_{\bs|\bw}, p_{\bx\bu|\bs\bw}^*, \scrW_K) \} .
\label{eq:I-sp-simple}
\end{equation}

{\bf (i). False Positives.}
From (\ref{eq:decision-simple}), the occurrence of a false positive implies that
\begin{equation}
   \exists \lambda \in \scrP_{S|S^d W}^{[N]}, l, m \notin \calK \,:\quad I(\bu(l,m, \lambda);\by|\bs^d\bw)
		> \rho(\lambda) + R+\Delta .
\label{eq:FP-simple}
\end{equation}
By construction of the codebook, $\bu(l,m, \lambda)$ is independent of $\by$ for $m \notin \calK$.
For any given $\lambda$, there are at most $2^{N \rho(\lambda)}$
possible values for $l$ and $2^{NR} -K$ possible values for $m$ in (\ref{eq:FP-simple}).
Hence the probability of false positives, conditioned on the joint type class
$T_{\by(\bx\bu)_{\calK} \bs\bw}$, is
\begin{eqnarray}
   \lefteqn{P_{FP}(T_{\by(\bx\bu)_{\calK} \bs\bw},\scrW_K)} \nonumber\\
	& \le & \sum_{\lambda} (2^{NR}-K) \, 2^{N\rho(\lambda)}
			\,Pr[I(\bu(l,m, \lambda);\by|\bs^d\bw) > \rho(\lambda) + R+\Delta ] 	\nonumber \\
	& \stackrel{(a)}{\doteq} & \sum_{\lambda} 2^{N (R+\rho(\lambda))}
			\,2^{-N (R+\Delta+\rho(\lambda))} 								\nonumber \\
	& \stackrel{(b)}{\le} & (N+1)^{|\calS|\,L_w} \,2^{-N \Delta} 				\nonumber \\
	& \doteq & 2^{-N \Delta}
\label{eq:P-FP-simple}
\end{eqnarray}
where (a) is obtained by application of (\ref{eq:I-rc-simple}) with
$\nu = \rho(\lambda) + R+\Delta$,
and (b) because the number of conditional types $\lambda$ is at most $(N+1)^{|\calS|\,L_w}$.

Averaging over all type classes $T_{\by(\bx\bu)_{\calK}\bs\bw}$, we obtain
$P_{FP} \dotle 2^{-N \Delta}$, from which (\ref{eq:E-FP-simple}) follows.

{\bf (ii). Detect-One Error Criterion} (Miss All Colluders).
We first derive the error exponent for the event that the decoder misses
a specific colluder $m \in \calK$.
Any coalition $\hat{\calK}$ that contains $m$ fails the test (\ref{eq:decision-simple}),
i.e., for any such $\hat{\calK}$,
\begin{equation}
   \forall \lambda \in \scrP_{S|S^d W}^{[N]} \,:\quad \max_l I(\bu(l,m, \lambda);\by|\bs^d\bw)
	\le \rho(\lambda) + R+\Delta .
\label{eq:miss-K*-simple}
\end{equation}
This implies that
\begin{equation}
   I(\bu(l,m, p_{\bs|\bs^d\bw});\by|\bs^d\bw) \le \rho(p_{\bs|\bs^d\bw}) + R+\Delta
\label{eq:miss-K*-2-simple}
\end{equation}
where $l$ is the row index actually selected by the encoder, and 
$p_{\bs|\bs^d\bw}$ is the actual host sequence conditional type.
The probability of the miss-$m$ event, given the joint type
$p_{\bw}^* \,p_{\bs|\bw} \,p_{\bx\bu|\bs\bw}^*$,
is therefore upper-bounded by the probability of the event (\ref{eq:miss-K*-2-simple}):
\begin{eqnarray*}
  p_{miss-m}(p_{\bw}^*,p_{\bs|\bw},p_{\bx\bu|\bs\bw}^*,\scrW_K)
   & \le & Pr \left[ I(\bu(l,m, p_{\bs|\bs^d\bw});\by|\bs^d\bw) \le \rho(p_{\bs|\bs^d\bw}) + R+\Delta \right] \\
	& \stackrel{(a)}{\dotle} & 
		\,\exp_2 \left\{ -N \breve{E}_{psp,m,N}(R+\Delta,L_w,L_u,p_{\bw}^*,p_{\bs|\bw},
		p_{\bx\bu|\bs\bw}^*,\scrW_K) \right\} 
\end{eqnarray*}
where (a) follows from (\ref{eq:I-sp-simple}) with $\nu = R+\Delta$.

The miss-all event is the intersection of the miss-$m$ events over $m \in \calK$.
Its conditional probability is
\begin{eqnarray}
  \lefteqn{p_{miss-all}(p_{\bw}^*,p_{\bs|\bw},p_{\bx\bu|\bs\bw}^*,\scrW_K)} \nonumber \\
	& = & Pr \left[ \bigcap_{m \in \calK} \left\{
		\mathrm{miss~} m ~|~p_{\bw}^*,p_{\bs|\bw},p_{\bx\bu|\bs\bw}^* \right\} \right] \nonumber \\
    & \le & \min_{m \in \calK} p_{miss-m}(p_{\bw}^*,p_{\bs|\bw},p_{\bx\bu|\bs\bw}^*,\scrW_K) \nonumber \\
	& \doteq & \exp_2 \left\{ -N \max_{m\in\calK} \breve{E}_{psp,m,N}(R+\Delta,L_w,L_u,
		p_{\bw}^*,p_{\bs|\bw},p_{\bx\bu|\bs\bw}^*,\scrW_K) \right\} .
\label{eq:pmiss-all-simple}
\end{eqnarray}

Averaging over $\bS$, we obtain
\begin{eqnarray*}
   \lefteqn{p_{miss-all}(\scrW_K)} \\
	& \le & \sum_{p_{\bs|\bw}} \,Pr[T_{\bs|\bw}] 
			\,p_{miss-all}(p_{\bw}^*,p_{\bs|\bw},p_{\bx\bu|\bs\bw}^*,\scrW_K) \\
	& \doteq & \max_{p_{\bs|\bw}} \,Pr[T_{\bs|\bw}] 
			\,p_{miss-all}(p_{\bw}^*,p_{\bs|\bw},p_{\bx\bu|\bs\bw}^*,\scrW_K) \\
	& \stackrel{(a)}{\doteq} & \max_{p_{\bs|\bw}} \exp_2 \left\{ -N \left[ D(p_{\bs|\bw} \| p_S \,|\,p_{\bw}^*)
			+ \max_{m\in\calK} \breve{E}_{psp,m,N}(R+\Delta,L_w,L_u,p_{\bw}^*,p_{\bs|\bw},
			p_{\bx\bu|\bs\bw}^*,\scrW_K) \right] \right\} \\
	& \stackrel{(b)}{\doteq} & \exp_2 \left\{ -N \overline{E}_{psp,N}(R+\Delta,L_w,L_u,D_1,\scrW_K) \right\} \\
	& \stackrel{(c)}{\doteq} & \exp_2 \left\{ -N \overline{E}_{psp}(R+\Delta,L_w,L_u,D_1,\scrW_K) \right\}
\end{eqnarray*}
which establishes (\ref{eq:E-one-simple}).
Here (a) follows from (\ref{eq:Pr-psw}) and (\ref{eq:pmiss-all-simple}),
(b) from (\ref{eq:Emax-tilde-simple-N}) and (\ref{eq:Emax-simple-N}),
and (c) from (\ref{eq:lim-Emax-simple-N}).

{\bf (iii). Detect-All Error Criterion} (Miss Some Colluders).

The miss-some event is the union of the miss-$m$ events over $m \in \calK$.
Given the joint type $p_{\bw}^* \,p_{\bs|\bw}\,p_{\bx\bu|\bs\bw}^*$,
the probability of this event is
\begin{eqnarray}
  \lefteqn{p_{miss-some}(p_{\bw}^*,p_{\bs|\bw},p_{\bx\bu|\bs\bw}^*,\scrW_K)} \\
	& = & Pr \left[ \bigcup_{m \in \calK} \left\{
			\mathrm{miss~} m ~|~p_{\bw}^*,p_{\bs|\bw},p_{\bx\bu|\bs\bw}^* \right\} \right] \nonumber \\
    & \le & \sum_{m \in \calK} p_{miss-m}(p_{\bw}^*,p_{\bs|\bw},p_{\bx\bu|\bs\bw}^*,\scrW_K) \nonumber \\
	& \doteq & \max_{m \in \calK} \exp_2 \left\{ -N \breve{E}_{psp,m,N}
			(R+\Delta,L_w,L_u,p_{\bw}^*,p_{\bs|\bw},p_{\bx\bu|\bs\bw}^*,\scrW_K) \right\} \nonumber \\
	& = & \exp_2 \left\{ -N \min_{m \in \calK} \breve{E}_{psp,m,N}(R+\Delta,L_w,L_u,p_{\bw}^*,
			p_{\bs|\bw},p_{\bx\bu|\bs\bw}^*,\scrW_K) \right\} .
\label{eq:pmiss-some-simple}
\end{eqnarray}

Averaging over $\bS$, we obtain
\begin{eqnarray*}
   \lefteqn{p_{miss-some}(\scrW_K)} \\
	& \le & \sum_{p_{\bs|\bw}} \,Pr[T_{\bs|\bw}] \,p_{miss-some}(p_{\bw}^*,p_{\bs|\bw},p_{\bx\bu|\bs\bw}^*,\scrW_K) \\
	& \stackrel{(a)}{\doteq} & \max_{p_{\bs|\bw}}
		\,\exp_2 \left\{ -N \left[ D(p_{\bs|\bw} \| p_S \,|\,p_{\bw}^*)
		+ \min_{m \in \calK} \breve{E}_{psp,m,N}(R+\Delta,L_w,L_u,p_{\bw}^*,
		p_{\bs|\bw},p_{\bx\bu|\bs\bw}^*,\scrW_K) \right] \right\} \\
	& \stackrel{(b)}{\le} & \exp_2 \left\{ -N \underline{E}_{psp,N}(R+\Delta,D_1,\scrW_K) \right\} \\
	& \stackrel{(c)}{\doteq} & \exp_2 \left\{ -N \underline{E}_{psp}(R+\Delta,L_w,L_u,D_1,\scrW_K) \right\}
\end{eqnarray*}
which establishes (\ref{eq:E-all-simple}).
Here (a) follows from (\ref{eq:Pr-psw}) and (\ref{eq:pmiss-some-simple}),
(b) from (\ref{eq:Emin-simple-N}) and
(\ref{eq:Emin-tilde-simple-N}), and (c) from (\ref{eq:lim-Emin-simple-N}).

{\bf (iv). Fair Collusion Channels.}
The proof parallels that of \cite[Theorem~4.1(iv)]{Moulin08}, using the conditional divergence
$D(\tp_{Y(XU)_{\calK}|SW} \,\tp_{S|W} \| \tp_{Y|X_{\calK}} \,p_{XU|SW}^{\calK} \,p_S \,| p_W)$
in place of $D(\tp_{YX_{\calK}|W} \| \tp_{Y|X_{\calK}} \,p_{X|W}^{\calK} \,| p_W)$.

{\bf (v).} Immediate, because $\overline{E}_{psp} = \underline{E}_{psp}$ in this case.

{\bf (vi). Positive Error Exponents.}
From Part (v) above, we may restrict our attention to $\scrW_K = \scrW_K^{fair}$.
Consider any $\calW = \{1,\cdots,L_w\}$ and $p_W$ that is positive over its support set
(if it is not, reduce the value of $L_w$ accordingly.)
For any $m \in \calK$, the minimand in the expression (\ref{eq:Epsp-m-simple}) for
$\tilde{E}_{psp,m}(R,L_w,L_u,p_W,p_{XU|SW}$, $\scrW_K^{fair})$ is zero if and only if 
\[ \tp_{Y(XU)_{\calK}|SW} \,\tp_{S|W} = \tp_{Y|X_{\calK}} \,p_{XU|SW}^{\calK} \,p_S,
	\quad \mathrm{with~} \tp_{Y|X_{\calK}} \in \scrW_K^{fair} . \]
Such $(\tp_{Y(XU)_{\calK}|SW}, \tp_{S|W})$ is feasible for (\ref{eq:set-m-simple}) if and only if
$(p_{XU|SW},\tp_{Y|X_{\calK}})$ is such that $I(U_m;Y|S^d,W)$ $\le I(U_m;S|S^d,W) + R$.
It is not feasible, and thus a positive exponent $E^{one}$ is guaranteed,
if $R < I(U_1;Y|S^d,W) - I(U_1;S|S^d,W)$. The supremum of all such $R$ is given by
(\ref{eq:C1}) and is achieved by letting $\epsilon \to 0$, $\Delta \to 0$,
and $L_w, L_u \to \infty$.
\hfill $\Box$

\section{Proof of Theorem~\ref{thm:joint}}
\label{Sec:JointProof}
\setcounter{equation}{0}

We derive the error exponents for the M2PMI decision rule (\ref{eq:M2PMI}).
Define for all $\calA \subseteq \calK$
\begin{eqnarray}
  \scrP_{Y(XU)_{\calK}|SW}^{[N]}(p_{\bw}, p_{\bs|\bw}, p_{\bx\bu|\bs\bw}, \scrW_K, R,L_w,L_u, \calA)
	& = & \left\{ p_{\by(\bx\bu)_{\calK}|\bs\bw}
		\,:~p_{(\bx\bu)_{\calK}|\bs\bw} \in \calM(p_{\bx\bu|\bs\bw}),
		\right. \nonumber\\  
	& & p_{\by|\bx_{\calK}} \in \scrW_K, \nonumber \\
	& & \left. \oI(\bu_\calA;\by\bu_{\calK\setminus\calA}|\bs^d\bw)
		\le |\calA| (\rho(p_{\bs|\bs^d\bw})+R) \right\} \label{eq:set-A-N} \\
  \breve{E}_{psp,\calA,N}(R,L_w,L_u,p_{\bw}, p_{\bs|\bw}, p_{\bx\bu|\bs\bw},\scrW_K)
	& = & \min_{p_{\by(\bx\bu)_{\calK}|\bs\bw} \,\in\,
		\scrP_{Y(XU)_{\calK}|SW}^{[N]}(p_{\bw}, p_{\bs|\bw}, p_{\bx\bu|\bs\bw}, \scrW_K, R,L_w,L_u, \calA)}
			\nonumber \\
	&   & \quad D(p_{\by(\bx\bu)_{\calK}|\bs\bw} \| p_{\by|\bx_{\calK}}
			\,p_{\bx\bu|\bs\bw}^{\calK} \,|\, p_{\bs\bw}) ,			\label{eq:Epsp-cond-N} \\
  \hat{E}_{psp,\calA,N}(R,L_w,L_u,p_{\bw}, p_{\bs|\bw}, p_{\bx\bu|\bs\bw},\scrW_K)
	& = & D(p_{\bs|\bw}\|p_S\,|p_{\bw}) + \breve{E}_{psp,\calA,N}(R,L_w,L_u, \nonumber \\
	& & \qquad p_{\bw}, p_{\bs|\bw}, p_{\bx\bu|\bs\bw},\scrW_K) \nonumber \\
	& = & \min_{p_{\by(\bx\bu)_{\calK}|\bs\bw} \,\in\,
		\scrP_{Y(XU)_{\calK}|SW}^{[N]}(p_{\bw}, p_{\bs|\bw}, p_{\bx\bu|\bs\bw}, \scrW_K, R,L_w,L_u, \calA)}
			\nonumber \\
	&   & \quad D(p_{\by(\bx\bu)_{\calK}|\bs\bw} \,p_{\bs|\bw} \| p_{\by|\bx_{\calK}}
			\,p_{\bx\bu|\bs\bw}^{\calK} \,p_S \,|\, p_{\bw}) ,			\label{eq:Epsp-A-N} \\
	\overline{\hat{E}}_{psp,N}(R,L_w,L_u,p_{\bw}, p_{\bs|\bw}, p_{\bx\bu|\bs\bw},\scrW_K)
	& = & \hat{E}_{psp,\calK,N}(R,L_w,L_u,p_{\bw}, p_{\bs|\bw}, p_{\bx\bu|\bs\bw},\scrW_K) ,
																		\label{eq:Emax-tilde-N} \\
   \underline{\hat{E}}_{psp,N}(R,L_w,L_u,p_{\bw}, p_{\bs|\bw}, p_{\bx\bu|\bs\bw},\scrW_K)
	& = & \min_{\calA \subseteq \calK}
			\hat{E}_{psp,\calA,N}(R,L_w,L_u,p_{\bw}, p_{\bs|\bw}, p_{\bx\bu|\bs\bw},\scrW_K) ,
																		\nonumber \\
																		\label{eq:Emin-tilde-N} \\
   E_{psp,N}(R,L_w,L_u,D_1,\scrW_K) & = & \max_{p_{\bw} \in \scrP_W^{[N]}}
		\;\min_{p_{\bs|\bw} \in \scrP_{S|W}^{[N]}}
		\max_{p_{\bx\bu|\bs\bw}  \in \scrP_{XU|SW}^{[N]}(p_{\bw} \,p_{\bs|\bw},L_w,L_u,D_1)} \nonumber \\
	& & \hat{E}_{psp,\calK,N}(R,L_w,L_u,p_{\bw}, p_{\bs|\bw}, p_{\bx\bu|\bs\bw},
		\scrW_{K_{nom}}^{fair}) .										\nonumber\\
																		\label{eq:Epsp-N}
\end{eqnarray}

Denote by $p_{\bw}^*$ and $p_{\bx\bu|\bs\bw}^*$ the maximizers in (\ref{eq:Epsp-N}),
the latter viewed as a function of $p_{\bs|\bw}$. Both maximizers depend implicitly
on $R$, $D_1$, and $\scrW_{K_{nom}}^{fair}$. Let
\begin{eqnarray}
   \overline{E}_{psp,N}(R,L_w,L_u,D_1,\scrW_K)
	& = & \min_{p_{\bs|\bw}} \overline{\hat{E}}_{psp,N}
			(R,L_w,L_u, p_{\bw}^*, p_{\bs|\bw}, p_{\bx\bu|\bs\bw}^*,\scrW_K)		\label{eq:Emax-N} \\
   \underline{E}_{psp,N}(R,L_w,L_u,D_1,\scrW_K)
	& = & \min_{p_{\bs|\bw}} \underline{\hat{E}}_{psp,N}
			(R,L_w,L_u, p_{\bw}^*, p_{\bs|\bw}, p_{\bx\bu|\bs\bw}^*,\scrW_K) .		\label{eq:Emin-N}
\end{eqnarray}
The exponents (\ref{eq:Epsp-A-N})---(\ref{eq:Emin-N}) differ from (\ref{eq:Epsp-A})---(\ref{eq:Emin})
in that the optimizations are performed over conditional types instead of general conditional
p.m.f.'s. We have
\begin{eqnarray}
 	\lim_{N \to \infty} \overline{E}_{psp,N}(R,L_w,L_u,D_1,\scrW_K)
		& = & \overline{E}_{psp}(R,L_w,L_u,D_1,\scrW_K)
														\label{eq:lim-Emax-N} \\
 	\lim_{N \to \infty} \underline{E}_{psp,N}(R,L_w,L_u,D_1,\scrW_K)
		& = & \underline{E}_{psp}(R,L_w,L_u,D_1,\scrW_K)
														\label{eq:lim-Emin-N}
\end{eqnarray}
by continuity of the divergence and mutual-information functionals.

The codebook and encoding procedure are exactly as in the proof of Theorem~\ref{sec:simple},
the difference being that $p_{\bw}^*$ and $p_{\bx\bu|\bs\bw}^*$ are solutions to
the optimization problem (\ref{eq:Epsp-N}) instead of (\ref{eq:Epsp-simple-N}).
The decoding rule is the M2PMI rule of (\ref{eq:M2PMI}).

To analyze the error probability for this random-coding scheme, it is again
sufficient to restrict our attention to strongly-exchangeable channels and
use the bound (\ref{eq:bound2}) on the conditional probability of the collusion
channel output. We also use Lemma~\ref{lem:3properties}.

{\bf (i). False Positives.}
By application of (\ref{eq:property1}),
a false positive occurs if $\hat{\calK} \setminus \calK \ne \emptyset$ and
\begin{eqnarray}
   \exists \lambda \in \scrP_{S|S^d W}^{[N]} \,:\quad \forall \calA \subseteq \hat{\calK} ~:\quad
  \exists l_{\hat{\calK}} :~
	& & \oI(\bu(l_{\calA}, m_{\calA}, \lambda);
		\by\bu(l_{\hat{\calK}\setminus\calA}, m_{\hat{\calK}\setminus\calA}, \lambda)
		\,|\bs^d\bw) \nonumber\\
	& & \qquad > |\calA| \,(\rho(\lambda) + R+\Delta) .
\label{eq:FP-condition0}
\end{eqnarray}
The probability of this event is upper-bounded by the probability of the larger event
\begin{eqnarray}
   \forall \calA \subseteq \hat{\calK} ~:\quad
  \exists \lambda, l_{\hat{\calK}} :~ 
	& & \oI(\bu(l_{\calA}, m_{\calA}, \lambda);
		\by\bu(l_{\hat{\calK}\setminus\calA}, m_{\hat{\calK}\setminus\calA},
		\lambda)\,|\bs^d\bw) \nonumber \\
	& & \qquad > |\calA| \,(\rho(\lambda) + R+\Delta) .
\label{eq:FP-condition1}
\end{eqnarray}
Denote by $p_{\bs|\bs^d\bw}^*$ the conditional type of the host sequence and by $l_{\calK}^*$
the row indices selected by the encoder.
To each triple $(\hat{\calK}, \lambda, l_{\hat{\calK}})$,
we associate a unique subset $\calB$ of $\calK \cap \hat{\calK}$ defined as follows:
\begin{itemize}
\item If $\lambda \ne p_{\bs|\bs^d\bw}^*$ then $\calB = \emptyset$
\item If $\lambda = p_{\bs|\bs^d\bw}^*$ then $\calB$ is the (possibly empty) set of all indices
	$k \in \calK \cap \hat{\calK}$ such that $l_k = l_k^*$.
\end{itemize}
Thus $\calB$ is the set of colluder indices $k\in\calK$ for which the decoder correctly
identifies the conditional host sequence type $p_{\bs|\bs^d\bw}^*$ and the codewords
$\bu(l_k^*,k,p_{\bs|\bs^d\bw}^*)$ that were assigned by the encoder.
Denoting by $\Omega(\calB)$ the set of pairs $(\lambda, l_{\hat{\calK}})$
associated with $\calB$, we rewrite (\ref{eq:FP-condition1}) as
\begin{eqnarray}
   \lefteqn{\forall \calA \subseteq \hat{\calK} ~:\quad
		\exists \calB \subseteq \calK \cap \hat{\calK}, 
		\;\exists (\lambda, l_{\hat{\calK}}) \in \Omega(\calB) :} \nonumber \\
   & & \oI(\bu(l_{\calA}, m_{\calA}, \lambda);
		\by\bu(l_{\hat{\calK}\setminus\calA}, m_{\hat{\calK}\setminus\calA}, \lambda)\,|\bs^d\bw)
		> |\calA| (\rho(\lambda) + R+\Delta) .
\label{eq:FP-condition2}
\end{eqnarray}

Define the complement set $\calA = \hat{\calK} \setminus \calB$ which is comprised of all
incorrectly accused users as well as any colluder $k$ such that $\lambda \ne p_{\bs|\bs^d\bw}^*$
or $l_k \ne l_k^*$. Since $\calB \subseteq \hat{\calK}$ and there is at least one innocent
user in $\hat{\calK}$, the cardinality of $\calA $ is at least equal to 1. 
By construction of the codebook and definition of $\calA$ and $\calB$,
$\bu(l_{\calA}, m_{\calA}, \lambda)$ is independent of
$\by$ and $\bu(l_{\calB}^*, m_{\calB}, p_{\bs|\bs^d\bw}^*)$.
The probability of the event (\ref{eq:FP-condition2}) is upper-bounded by the probability
of the larger event
\begin{equation}
   \exists \calB \subseteq \calK, 
		\;\exists \lambda, l_{\calA}, m_{\calA} \,:\quad
   \oI(\bu(l_{\calA}, m_{\calA}, \lambda); \by\bu(l_{\calB}^*, m_{\calB},
		p_{\bs|\bs^d\bw}^*)\,|\bs^d\bw) > |\calA| (\rho(\lambda) + R+\Delta) .
\label{eq:event-FP}
\end{equation}

Hence the probability of false positives, conditioned on $T_{\by(\bx\bu)_{\calK}\bs\bw}$, satisfies
\begin{eqnarray}
   \lefteqn{P_{FP}(T_{\by(\bx\bu)_{\calK}\bs\bw},\scrW_K)} \nonumber \\
	& = & Pr \left[ \bigcup_{\calB \subseteq \calK} \bigcup_{|\calA| \ge 1} \left\{
		\exists \lambda, l_{\calA}, m_{\calA} \,:
		\oI(\bu(l_{\calA}, m_{\calA}, \lambda); \by\bu(l_{\calB}^*, m_{\calB},
		p_{\bs|\bs^d\bw}^*)\,|\bs^d\bw) \right. \right. \nonumber \\
	& & \hspace*{2.5in} \left. \left. > |\calA| \,(\rho(\lambda) + R+\Delta)
		\right\} \right] \nonumber \\
	& \le & \sum_{\calB \subseteq \calK} 
		\sum_{|\calA| \ge 1} P_{\calB,|\calA|}(T_{\by(\bx\bu)_{\calK}\bs\bw},\scrW_K)
\label{eq:P-FP1}
\end{eqnarray}
where
\begin{eqnarray}
   P_{\calB,|\calA|}(T_{\by(\bx\bu)_{\calK}\bs\bw},\scrW_K)
		= Pr \left[ \exists \lambda, l_{\calA}, m_{\calA} \right. :
	& & \oI(\bu(l_{\calA}, m_{\calA}, \lambda); \by\bu(l_{\calB}^*, m_{\calB},
		p_{\bs|\bs^d\bw}^*)\,|\bs^d\bw) \nonumber \\
	& & \qquad \left. > |\calA| \,(\rho(\lambda) + R+\Delta) \right] .
\label{eq:FP1+2}
\end{eqnarray}
By definition of $\calB$, there are at most $\sum_{\lambda \ne p_{\bs|\bs^d\bw}}
2^{N |\calA| \,\rho(\lambda)}$ possible values for $l_{\calA}$ and
$2^{N |\calA| R}$ possible values for $m_{\calA}$ in (\ref{eq:FP1+2}). Hence
\begin{eqnarray}
   \lefteqn{P_{\calB,|\calA|}(T_{\by(\bx\bu)_{\calK}\bs\bw},\scrW_K)} \nonumber \\
	& \le & \sum_{\lambda} 2^{N |\calA| (R+\rho(\lambda))}
			\,Pr[\oI(\bu(l_{\calA}, m_{\calA}, \lambda);
			\by\bu(l_{\calB}^*, m_{\calB}, p_{\bs|\bs^d\bw}^*)\,|\bs^d\bw)	
			> |\calA| \,(\rho(\lambda) + R+\Delta) ] 							\nonumber \\
	& \stackrel{(a)}{\doteq} & \sum_{\lambda} 2^{N |\calA| (R+\rho(\lambda))}
			\,2^{-N |\calA| \,(R+\Delta+\rho(\lambda))} 					\nonumber \\
	& \le & (N+1)^{|\calS|} \,2^{-N |\calA| \Delta} 					\nonumber \\
	& \doteq & 2^{-N |\calA| \Delta}
\label{eq:P-FP2}
\end{eqnarray}
where (a) is obtained by application of (\ref{eq:I-rc}) with
$\by\bu(l_{\calB}^*, m_{\calB}, p_{\bs|\bs^d\bw}^*)$ in place of $\bz$.

Combining (\ref{eq:P-FP1}) and (\ref{eq:P-FP2}) we obtain
\begin{eqnarray*}
   P_{FP}(T_{\by(\bx\bu)_{\calK}\bs\bw},\scrW_K)
	& \le & \sum_{\calB \subseteq \calK} \sum_{|\calA| \ge 1} 2^{-N |\calA| \Delta} \\
	& \doteq & 2^{-N\Delta} .
\end{eqnarray*}
Averaging over all joint type classes $T_{\by(\bx\bu)_{\calK}\bs\bw}$,
we obtain $P_{FP} \dotle 2^{-N \Delta}$, from which (\ref{eq:E-FP}) follows.

\newpage

{\bf (ii). Detect-All Criterion.} (Miss Some Colluders.)

Under the miss-some error event,
{\em any} coalition $\hat{\calK}$ that {\em contains} $\calK$ fails the test.
By (\ref{eq:property1}), this implies
\begin{eqnarray}
   \forall \lambda \in \scrP_{S|S^d W}^{[N]} \,:\quad \exists \calA \subseteq \hat{\calK} ~:
	& & \max_{l_{\hat{\calK}}} \oI(\bu(l_{\calA}, m_{\calA}, \lambda);
		\by\bu(l_{\hat{\calK}\setminus\calA}, m_{\hat{\calK}\setminus\calA},
		\lambda)\,|\bs^d\bw) \nonumber \\
	& & \qquad \le  |\calA| \,(\rho(\lambda) + R+\Delta) .
\label{eq:miss-K*}
\end{eqnarray}
In particular, for $\hat{\calK} = \calK$ we have
\begin{equation}
   \exists \calA \subseteq \calK ~:\quad \oI(\bu(l_{\calA}, m_{\calA}, p_{\bs|\bs^d\bw});
		\by\bu(l_{\calK\setminus\calA}, m_{\calK\setminus\calA}, p_{\bs|\bs^d\bw})\,|\bs^d\bw)
		\le |\calA| \,(\rho(p_{\bs|\bs^d\bw}) + R+\Delta) .
\label{eq:miss-K*-2}
\end{equation}
where $l_{\calK}$ are the row indices actually selected by the encoder, and 
$p_{\bs|\bs^d\bw}$ is the actual host sequence conditional type.
The probability of the miss-some event, conditioned on $(\bs,\bw)$,
is therefore upper bounded by the probability of the event (\ref{eq:miss-K*-2}):
\begin{eqnarray}
  \lefteqn{p_{miss-some}(p_{\bw}^*,p_{\bs|\bw},p_{\bx\bu|\bs\bw}^*,\scrW_K)} \nonumber \\
	& \le & Pr \left[ \bigcup_{\calA \subseteq \calK} \left\{
		\oI(\bu(l_{\calA}, m_{\calA}, p_{\bs|\bs^d\bw});
		\by\bu(l_{\calK\setminus\calA}, m_{\calK\setminus\calA}, p_{\bs|\bs^d\bw})\,|\bs^d\bw)
		\le |\calA| (\rho(p_{\bs|\bs^d\bw}) + R+\Delta) \right\} \right] \nonumber \\
   & \le & \sum_{\calA \subseteq \calK} Pr \left[ \oI(\bu(l_{\calA}, m_{\calA}, p_{\bs|\bs^d\bw});
		\by\bu(l_{\calK\setminus\calA}, m_{\calK\setminus\calA}, p_{\bs|\bs^d\bw})\,|\bs^d\bw)
		\le |\calA| (\rho(p_{\bs|\bs^d\bw}) + R+\Delta) \right] \nonumber \\
	& \stackrel{(a)}{\dotle} & \sum_{\calA \subseteq \calK} \exp_2 \left\{
		-N \breve{E}_{psp,\calA,N}(R+\Delta,L_w,L_u,p_{\bw}^*,p_{\bs|\bw},p_{\bx\bu|\bs\bw}^*,\scrW_K) \right\}
																			\nonumber \\
	& \doteq & \max_{\calA \subseteq \calK} \exp_2 \left\{
		-N \breve{E}_{psp,\calA,N}(R+\Delta,L_w,L_u,p_{\bw}^*,p_{\bs|\bw},p_{\bx\bu|\bs\bw}^*,\scrW_K) \right\}
																			\nonumber \\
	& = & \exp_2 \left\{ - N \min_{\calA \subseteq \calK} 
		\breve{E}_{psp,\calA,N}(R+\Delta,L_w,L_u,p_{\bw}^*,p_{\bs|\bw},p_{\bx\bu|\bs\bw}^*,\scrW_K) \right\}
\label{eq:miss-K*-UB}
\end{eqnarray}
where (a) follows from (\ref{eq:I-sp}) with $\nu = R+\Delta$.

Averaging over $\bS$, we obtain
\begin{eqnarray*}
   \lefteqn{p_{miss-some(\scrW_K)}} \\
	& = & \sum_{p_{\bs|\bw}} \,Pr[T_{\bs|\bw}] \,p_{miss-some}(p_{\bw}^*,p_{\bs|\bw},p_{\bx\bu|\bs\bw}^*,\scrW_K) \\
	& \stackrel{(a)}{\doteq} & \max_{p_{\bs|\bw}}
		\exp_2 \left\{ -N [D(p_{\bs|\bw}\|p_S \,|\,p_{\bw}) + \min_{\calA \subseteq \calK} \,\breve{E}_{psp,N}
			(R+\Delta,L,p_{\bw}^*,p_{\bs|\bw},p_{\bx\bu|\bs\bw}^*,\scrW_K)] \right\} \\
	& \stackrel{(b)}{=} & \max_{p_{\bs|\bw}} \exp_2 \left\{
			-N \underline{\hat{E}}_{psp,N}(R+\Delta,L_w,L_u,p_{\bw}^*,p_{\bs|\bw},p_{\bx\bu|\bs\bw}^*,
			\scrW_K) \right\} \\	
	& \stackrel{(c)}{=} & \exp_2 \left\{
			-N \underline{E}_{psp,N}(R+\Delta,L_w,L_u,D_1,\scrW_K) \right\} \\	
	& \stackrel{(d)}{\doteq} & \exp_2 \left\{
			-N \underline{E}_{psp}(R+\Delta,L_w,L_u,D_1,\scrW_K) \right\} 
\end{eqnarray*}
which proves (\ref{eq:E-all}). Here (a) follows from (\ref{eq:Pr-psw}) and (\ref{eq:miss-K*-UB}),
(b) from the definitions (\ref{eq:Emin-tilde-N}) and (\ref{eq:Epsp-A-N}),
(c) from (\ref{eq:Emin-N}),
and (d) from the limit property (\ref{eq:lim-Emin-N}).

{\bf (iii). Detect-One Criterion} (Miss All Colluders.)
Either the estimated coalition $\hat{\calK}$ is empty, or it is
a set $\calI$ of innocent users (disjoint with $\calK$). Hence
$P_e^{one} \le Pr[\hat{\calK} = \emptyset] + Pr[\hat{\calK} = \calI]$.
The first probability, conditioned on $(\bs^d,\bw)$, is bounded as
\begin{eqnarray}
   Pr[\hat{\calK} = \emptyset]
	& = & Pr[\forall \calK' ~:~M2PMI(\calK') \le 0] \nonumber \\
	& \le & Pr[M2PMI(\calK) \le 0] \nonumber \\
	& = & Pr[\oI(\bu_{\calK};\by|\bs^d\bw) \le K(\rho(p_{\bs|\bs^d\bw})+R+\Delta)] \label{eq:empty-Khat} \\
	& \stackrel{(a)}{\doteq} & \exp_2 \left\{ -N \breve{E}_{psp,\calK,N}
			(R+\Delta,L_w,L_u,p_{\bw}^*,p_{\bs|\bw},p_{\bx\bu|\bs\bw}^*,\scrW_K) \right\} . \nonumber
\end{eqnarray}
where (a) follows from (\ref{eq:I-sp}) with $\nu = R+\Delta$.
To bound $Pr[\hat{\calK} = \calI]$, we use property (\ref{eq:property2})
with $\hat{\calK} = \calI$ and $\calA=\calK$, which yields
\[ \oI(\bu_{\calK};\by\bu_{\calI}|\bs^d\bw) \le K(\rho(p_{\bs|\bs^d\bw})+R+\Delta) . \]
Since
\[ \oI(\bu_{\calK};\by\bu_{\calI}|\bs^d\bw)
	= \oI(\bu_{\calK};\by|\bs^d\bw) + I(\bu_{\calK};\bu_{\calI}|\by\bs^d\bw)
	\ge \oI(\bu_{\calK};\by|\bs^d\bw) \]
combining the two inequalities above yields
\[ \oI(\bu_{\calK};\by|\bs^d\bw) \le K(\rho(p_{\bs|\bs^d\bw})+R+\Delta) . \]
The probability of this event is again given by (\ref{eq:empty-Khat});
we conclude that
\[ p_{miss-all}(p_{\bw}^* \,p_{\bs|\bw},p_{\bx\bu|\bs\bw}^*,\scrW_K)
	\doteq \exp_2 \left\{ -N \breve{E}_{psp,\calK,N}
			(R+\Delta,L_w,L_u,p_{\bw}^*,p_{\bs|\bw},p_{\bx\bu|\bs\bw}^*,\scrW_K) \right\} .
\]

Averaging over $\bS$ and proceeding as in Part (ii) above, we obtain
\begin{eqnarray*}
   p_{miss-all}(\scrW_K)
	& \le & \sum_{p_{\bs|\bw}} \,Pr[T_{\bs|\bw}]
		\,p_{miss-all}(p_{\bw}^* \,p_{\bs|\bw},p_{\bx\bu|\bs\bw}^*,\scrW_K) \\
	& \doteq & \exp_2 \left\{ -N \overline{E}_{psp}(R+\Delta,L_w,L_u,D_1,\scrW_K) \right\}
\end{eqnarray*}
which establishes (\ref{eq:E-one}).

{\bf (iv). Optimal Collusion Channels are Fair.}
The proof parallels that of \cite[Theorem~4.1(iv)]{Moulin08} and is omitted.

{\bf (v). Detect-All Exponent for Fair Collusion Channels.}
The proof parallels that of \cite[Theorem~4.1(v)]{Moulin08} and is omitted.

{\bf (vi). Achievable Rates.}
Consider any $\calW = \{1,\cdots,L_w\}$ and $p_W$ that is positive over its support set
(if it is not, reduce the value of $L_w$ accordingly.)
For any $\calA \subseteq \calK$, the divergence to be minimized in the expression
(\ref{eq:Epsp-A}) for
$\tilde{E}_{psp,\calA}(R,L_w,L_u,p_W,\tp_{S|W},p_{XU|SW},\scrW_K)$ is zero if and only if 
\[ \tp_{Y(XU)_{\calK}|SW} = \tp_{Y|X_{\calK}} \,p_{XU|SW}^{\calK} 
	\quad \mathrm{and~} \tp_{S|W} = p_S . \]
These p.m.f.'s are feasible for (\ref{eq:set-A}) if and only if 
the inequality below holds:
\[ \frac{1}{|\calA|} I(U_{\calA};YU_{\calK\setminus\calA}|S^d,W) > I(U;S|S^d,W) + R . \]
They are infeasible, and thus positive error exponents are guaranteed, if
\[ R < \min_{\calA \subseteq \calK}
	\frac{1}{|\calA|} I(U_{\calA};YU_{\calK\setminus\calA}|S^d,W) - I(U;S|S^d,W) .\]

From Part (iv) above, we may restrict our attention to $\scrW_K = \scrW_K^{fair}$
under the detect-one criterion. Since the p.m.f. of $(S,W,(XU)_{\calK},Y)$ is permutation-invariant,
by application of \cite[Eqn.~(3.3)]{Moulin08} with $(U_{\calK}, S^d)$ in place of
$(X_{\calK}, S)$, we have
\begin{equation}
   \min_{\calA \subseteq \calK} \frac{1}{|\calA|} I(U_{\calA};YU_{\calK\setminus\calA}|S^d W)
	 = \frac{1}{K} I(U_{\calK};Y|S^d W) .
\label{eq:I-min}
\end{equation}
Hence the supremum of all $R$ for error exponents are positive is given by
$\underline{C}^{one}(D_1,\scrW_K)$ in (\ref{eq:C-achieve-one}) and is obtained by letting
$\epsilon \to 0$, $\Delta \to 0$ and $L_w, L_u \to \infty$.

For any $\scrW_K$, under the detect-all criterion,
the supremum of all $R$ for which error exponents are positive
is given by $\underline{C}^{all}(D_1,\scrW_K)$ in (\ref{eq:C-achieve-all}) and is obtained
by letting $\epsilon \to 0$, $\Delta \to 0$ and $L_w,L_u \to \infty$. Since the optimal
conditional p.m.f. is not necessarily permutation-invariant, (\ref{eq:I-min}) does not hold
in general. However, if $\scrW_K = \scrW_K^{fair}$, (\ref{eq:I-min}) holds, and
the same achievable rate is obtained for the detect-one and detect-all problems.
\hfill $\Box$

\section{}
\label{Sec:lemma}
\setcounter{equation}{0}

\begin{lemma}
\noindent
1) Fix $(\bs^d,\bw)$ and $\bz \in \calZ^N$, and draw $\bu_{\calK} = \{\bu_m, \,m \in \calK\}$
i.i.d. uniformly over a common type class $T_{\bu|\bs^d\bw}$, independently of $\bz$.
We have the asymptotic equality
\begin{equation}
   Pr[T_{\bu_{\calK}|\bz\bs^d\bw}] = \frac{|T_{\bu_{\calK}|\bz\bs^d\bw}|}{|T_{\bu|\bs^d\bw}|^K}
	\doteq 2^{-N [KH(\bu|\bs^d\bw)-H(\bu_{\calK}|\bz\bs^d\bw)]} = 2^{-N \oI(\bu_{\calK};\bz|\bs^d\bw)}
\label{eq:cond-type-prob}
\end{equation}
\begin{eqnarray}
   Pr[\oI(\bu_{\calK};\bz|\bs^d\bw) \ge \nu] & \doteq & 2^{-N \nu} .
\label{eq:I-rc}
\end{eqnarray}
2) Given $\bw$, draw $\bs$ i.i.d. $p_S$. We have \cite{Csiszar81}
\begin{equation}
  Pr[T_{\bs|\bw}] \doteq 2^{-N D(p_{\bs|\bw}\|p_S|p_{\bw})} .
\label{eq:Pr-psw}
\end{equation}
3) Given $(\bs,\bw)$, draw $(\bx_k, \bu_k), \,k\in\calK$, i.i.d. uniformly from a conditional type
class $T_{\bx\bu|\bs\bw}$, and then draw $\bY$ uniformly from
a single conditional type class $T_{\by|\bx_{\calK}}$. We have
\begin{eqnarray}
   Pr[T_{\by(\bx\bu)_{\calK}|\bs\bw}]
	& = & \frac{|T_{\by|(\bx\bu)_{\calK} \bs\bw}|}{|T_{\by|\bx_{\calK}}|}
			\,\frac{|T_{(\bx\bu)_{\calK}|\bs\bw}|}{|T_{\bx\bu|\bs\bw}|^K} \nonumber \\
	& \doteq & \exp_2 \left\{ -N D(p_{\by\bx_{\calK}|\bs\bw} \|
			p_{\by|\bx_{\calK}} \,p_{\bx\bu|\bs\bw}^K \,|\,p_{\bs\bw}) \right\} .
\label{eq:PrT}
\end{eqnarray}

For any feasible, strongly exchangeable collusion channel, for any
$\calA \subseteq \calK$ and $\nu > 0$, we have
\begin{eqnarray}
   \lefteqn{Pr[\oI(\bu_{\calA};\by\bu_{\calK\setminus\calA}|\bs^d\bw)
				\le |\calA| (\nu + \rho(p_{\bs|\bs^d\bw}))]} \nonumber \\
	& \doteq & \exp_2 \left\{ -N \breve{E}_{psp,\calA,N}(\nu,L,p_{\bw}^*,p_{\bs|\bw},
				p_{\bx\bu|\bs\bw}^*,\scrW_K) \right\} .
\label{eq:I-sp}
\end{eqnarray}
\label{lem:3properties}
\end{lemma}
{\em Proof:}
The derivation of (\ref{eq:PrT}), (\ref{eq:Pr-psw}), and (\ref{eq:I-sp}) parallels
that of (D.12), (D.15) and (D.16) in \cite{Moulin08}.

\section{Proof of Theorem \ref{thm:public}}
\label{Sec:ConverProof}
\setcounter{equation}{0}

Let $K$ be size of the coalition and $(f_N,g_N)$ a sequence of length-$N$,
rate-$R$ randomized codes. We show that for any sequence of such
codes, reliable decoding of all $K$ fingerprints is possible only if
$R \le \overline{C}^{all}(D_1,\scrW_K)$.
Recall that the encoder generates marked copies $\bx_m = f_N(\bs,v,m)$ for
$1 \le m \le 2^{NR}$ and that the decoder outputs an estimated coalition
$g_N(\by,\bs^d,v) \in \{ 1, \cdots, 2^{NR} \}^\star$.
We use the notation $M^K \triangleq \{ M_1, \cdots, M_K\}$ and
$\bX^K \triangleq \{ \bX_1, \cdots, \bX_K\}$.

To prove that $\overline{C}^{all}(D_1,\scrW_K)$ is an upper bound on capacity,
it suffices to identify a family of collusion channels
for which reliable decoding is impossible at rates above
$\overline{C}^{all}(D_1,\scrW_K)$. As shown in \cite{Moulin08}, it is sufficient
to derive such a bound for the compound family $\scrW_K$ of {\em memoryless channels}.

Our derivation is an extension of the single-user
compound Gel'fand-Pinsker problem \cite{Moulin07} to the multiple-access case.
A lower bound on error probability is obtained when an oracle informs the decoder
that the coalition size is {\em at most} $K$.

There are $\left( \begin{array}{c} 2^{NR} \\ K \end{array} \right) \le 2^{KNR}$
possible coalitions of size $\le K$.
We represent such a coalition as $M^K \triangleq \{ M_1, \cdots, M_K\}$, where
$M_k, \,1 \le k \le K$, are drawn i.i.d. uniformly from $\{1, \cdots, 2^{NR}\}$.

Given a memoryless channel $p_{Y|X^K} \in \scrW_K$,
the joint p.m.f. of $(M^K,V,\bS,\bX^K,\bY)$ is given by
\begin{equation}
 p_{M^K V \bS \bX^K \bY} = p_S^N \,p_V \,\prod_{1 \le k \le K} \left(
	p_{M_k}\,\mathds 1_{\{\bX_k=f_N(\bS,V,M_k)\}} \right) p_{Y|X^K}^N .
\label{eq:jointProb}
\end{equation}

Our derivations make repeated use of the identity
\[ I(U_{\calA};Y|Z,U_{\calK\setminus\calA}) - I(U_{\calA};S|Z,U_{\calK\setminus\calA})
	= I(U_{\calA};Y,Z|U_{\calK\setminus\calA}) - I(U_{\calA};S,Z|U_{\calK\setminus\calA})
\]
which follows from the chain rule for conditional mutual information
and holds for any $(U_{\calK},S,Y,Z)$.

The total error probability (including false positives and false negatives)
for the detect-all decoder is
\begin{equation}
   P_e(p_{Y|X^K}) = Pr[\hat{\calK} \ne \calK]
\label{eq:Pe-1}
\end{equation}
when collusion channel $p_{Y|X^K} \in \scrW_K$ is in effect.

\underline{\em Step~1.}
Following the derivation of \cite[Eqn.~(B.20)]{Moulin08} with $(\bY,\bS^d,V)$
in place of $(\bY,\bS,V)$ at the receiver, for the error probability $P_e(p_{Y|X^K})$
to vanish for each $p_{Y|X^K} \in \scrW_K$, we need
\begin{equation}
   R \le \liminf_{N \to \infty} \min_{p_{Y|X^K} \in \scrW_K}
	\min_{\calA \subseteq \calK} \;\frac{1}{N |\calA|} I(M_{\calA};\bY|\bS^d,V).
\label{eq:KR}
\end{equation}

\underline{\em Step~2.}
Define the i.i.d. random variables
\begin{equation}
  W_i = \{V,\;S_j, j \ne i\} \in \calV_N \times \calS^{N-1} , \quad 1 \le i \le N .
\label{eq:Wi}
\end{equation}
Also define the random variables
\begin{eqnarray}
   V_{ki} & = & (M_k,V,S_{i+1}^N ),  \nonumber\\
   U_{ki} & = & (V_{ki},(YS^d)^{i-1} ) = (M_k,V,S_{i+1}^N,(YS^d)^{i-1} ),
					\quad 1 \le k \le K, \;1 \le i \le N
\label{eq:VW}
\end{eqnarray}
where $S_{i+1}^N \triangleq (S_{i+1}, \cdots, S_N)$ and
$(YS^d)^{i-1} \triangleq (Y_1, S_1^d,\cdots,Y_{i-1},S_{i-1}^d)$. Hence
\begin{equation}
   V_{i-1}^K = (V_i^K,S_i), \quad V_1^K = U_1^K, \quad V_N^K = (M^K, V) .
\label{eq:VW-recursive}
\end{equation}
The following properties hold for each $1 \le i \le N$:
\begin{itemize}
\item By (\ref{eq:jointProb}) and (\ref{eq:VW}),
	$(S_i,W_i,U_i^K) = (M^K,V,\bS,Y^{i-1}) \to X_i^K \to Y_i$ forms a Markov chain.
\item The random variables $X_{ki}, \,1 \le k \le K$, are conditionally i.i.d.
	given $(\bS,V)=(S_i,W_i)$.
\item Due to the term $Y^{i-1}$ in (\ref{eq:VW}), the random variables
	$U_{ki}, \,1 \le k \le K$, are conditionally {\em dependent} given
	$(\bS,V)=(S_i,W_i)$.
\end{itemize}
The joint p.m.f. of $(S_i,W_i,X_i^K,U_i^K,Y_i)$ may thus be written as
\begin{equation}
   p_{S_i} p_{W_i} \left( \prod_{1 \le k \le K} p_{X_{ki}|S_i W_i} \right)
	\,p_{U_i^K|X_i^K S_i W_i} \;p_{Y|X_{\calK}} , \quad 1 \le i \le N .
\label{eq:JointProb-i}
\end{equation}

\underline{\em Step~3.}
Consider a time-sharing random variable $T$ that is uniformly distributed
over $\{1,\cdots,N\}$ and independent of the other random variables, and define
the tuple of random variables $(S, S^d, W, U^K, X^K, Y)$
as $(S_T, S_T^d, W_T, U_T^K, X_T^K, Y_T)$.
Also let $W = (W_T,T)$ and $U_k = (U_{k,T},T)$, $1 \le k \le K$,
which are defined over alphabets of respective cardinalities
\[ L_w(N) = N \,|\calV_N| \,|\calS|^{N-1} \]
and
\[ L_u(N) = N \,|\calV_N| \,2^{N\left[R+\log\max(|\calS|,|\calY|\,|\calS^d|)\right]} . \]
Since $(S_i,W_i,U_i^K) \to X_i^K \to Y_i$ forms a Markov chain, so does
$(S, W, U^K) \to X^K \to Y$. From (\ref{eq:JointProb-i}),
the joint p.m.f. of $(S, W, U^K, X^K, Y)$ takes the form
\begin{equation}
   p_S p_W \left( \prod_{1 \le k \le K} p_{X_k|SW} \right) \,p_{U^K|X^K SW} \;p_{Y|X_{\calK}} .
\label{eq:joint-p.m.f.}
\end{equation}
In (\ref{eq:PXUS-outer-set}) we have defined the set
\begin{eqnarray}
   \lefteqn{\scrP_{X^K U^K W|S}^{outer}(p_S,L_w,L_u,D_1) = \left\{ p_{X^K U^K W|S}
		= p_W \,\left( \prod_{k=1}^K p_{X_k|SW} \right) \,p_{U^K|X^KSW} \right. } \nonumber \\
	& & \hspace*{1.8in} \left. ~:~p_{X_1|SW} = \cdots = p_{X_K|SW},
		\;\;\mathrm{and}\;\;\eE d(S,X_1) \le D_1 \right\}
\label{eq:PXUQS-set}
\end{eqnarray}
where $|\calW| = L_w$ and $|\calU| = L_u$. Observe that $p_{X^K U^K W|S}$ defined
in (\ref{eq:joint-p.m.f.}) belongs to $\scrP_{X^K U^K W|S}(p_S,L_w$, $L_u,D_1)$.

Define the collection of $K$ indices $\calK = \{1,2,\cdots,K\}$
and the following functionals indexed by $\calA \subseteq \calK$:
\begin{eqnarray}
  J_{L_w,L_u,\calA}(p_S,p_{X^K U^K W|S}, p_{Y|X^K})
	& = & \frac{1}{|\calA|} [I(U_{\calA};YS^d|U_{\calK\setminus\calA})
			- I(U_{\calA};S|U_{\calK\setminus\calA})] .
\label{eq:JL}
\end{eqnarray}

\underline{\em Step~4.}
We have
\begin{eqnarray}
   I(M_\calK;\bY|\bS^d,V)
	& \stackrel{(a)}{=} & I(M_\calK;\bY|\bS^d,V) - I(M_\calK,V;\bS|\bS^d) \nonumber \\
	& = & I(M_\calK,V;\bY|\bS^d) - I(V;\bY|\bS^d) - I(M_\calK,V;\bS|\bS^d) \nonumber \\
	& \le & I(M_\calK,V;\bY|\bS^d) - I(M_\calK,V;\bS|\bS^d) \nonumber \\
	& \stackrel{(b)}{=} & I(M_\calK,V;\bY\bS^d) - I(M_\calK,V;\bS) \nonumber \\
	& \stackrel{(c)}{\le} & \sum_{i=1}^N [I(U_{\calK,i};Y_i S_i^d) - I(U_{\calK,i};S_i)] \nonumber \\
	&=& I(U_{\calK,T};YS^d|T)- I(U_{\calK,T};S|T) \nonumber \\
	&=& I(U_{\calK,T},T;YS^d)-I(T;YS^d)-I(U_{\calK,T},T;S)+I(T;S) \nonumber\\
	&\stackrel{(d)}{\le}& I(U_{\calK,T},T;YS^d)-I(U_{\calK,T},T;S) \nonumber\\
	&\stackrel{(e)}{=}& I(U_{\calK};YS^d)-I(U_{\calK};S) \nonumber\\
	&=& K \,J_{L_w(N),L_u(N),\calK}(p_S,p_{X^K U^K W|S}, p_{Y|X^K}) ,
\label{eq:IK}
\end{eqnarray}
where (a) holds because $M_K, V, \bS$ are mutually independent, and
(b) follows from the chain rule for mutual information,
(c) from \cite[Lemma~4]{Gelfand80},
using $V_i^K$ and $U_i^K$ in place of $V_i$ and $U_i$, respectively,
(d) holds because $I(T;S)=0$, and (e) by definition of $U_{\calK}$.

For all $\calA \subset \calK$, we have
\begin{eqnarray}
   I(M_\calA;\bY|\bS^d,V)
	& = & I(M_\calA,V;\bY|\bS^d,V) \nonumber \\
	& \stackrel{(a)}{=} & I(M_\calA,V;\bY|\bS^d,V) - I(M_\calA,V;\bS|\bS^d,M_{\calK\setminus\calA},V)
															\nonumber \\
	& \stackrel{(b)}{=} & I(M_\calA,V;\bY|\bS^d,M_{\calK\setminus\calA},V)
			- I(M_\calA,V;\bS|\bS^d,M_{\calK\setminus\calA}) \nonumber \\
	& = & I(M_\calA,V;\bY\bS^d|M_{\calK\setminus\calA},V)
			- I(M_\calA,V;\bS|\bS^d,M_{\calK\setminus\calA},V) \nonumber \\
	& \stackrel{(c)}{=} & \sum_{i=1}^N [I(U_{\calA,i};Y_i S_i^d|U_{\calK\setminus\calA,i})
			- I(U_{\calA,i};S_i|U_{\calK\setminus\calA,i})] \label{eq:I-GPineq-cond} \\
	& = & N \,[I(U_{\calA,T};Y S^d|U_{\calK\setminus\calA,T},T)
			- I(U_{\calA,T};S|U_{\calK\setminus\calA,T},T)] \nonumber \\
	& = & N \,[I(U_{\calA,T},T;Y S^d|U_{\calK\setminus\calA,T},T)
			- I(U_{\calA,T},T;S|U_{\calK\setminus\calA,T},T)] \nonumber \\
	& \stackrel{(d)}{=} & N \,[I(U_{\calA};Y S^d|U_{\calK\setminus\calA})
			- I(U_{\calA};S|U_{\calK\setminus\calA})] \nonumber \\
	& = & N \,|\calA|\,J_{L_w(N),L_u(N),\calA}(p_S,p_{X^K U^K W|S}, p_{Y|X^K}) .
\label{eq:IA}
\end{eqnarray}
where (a) and (b) hold because $M_K$, $\bS$, and $V$ are mutually independent, 
the equality (c) is proved at the end of this section,
and (d) follows from the definition of $U_\calK$.

Combining (\ref{eq:KR}), (\ref{eq:IK}), and (\ref{eq:IA}), we obtain
\begin{eqnarray}
R &\le& \liminf_{N \to \infty} \min_{p_{Y|X^K} \in \scrW_K}
	\min_{\calA \subseteq \calK} J_{L_w(N),L_u(N),\calA}(p_S,p_{X^K U^K W|S}, p_{Y|X^K}) \nonumber \\
&\stackrel{(a)}{\le}& \sup_{L_w,L_u} \min_{p_{Y|X^K} \in \scrW_K}
	\min_{\calA \subseteq \calK} J_{L_w,L_u,\calA}(p_S,p_{X^K U^K W|S}, p_{Y|X^K}) \nonumber \\
&\le& \sup_{L_w,L_u} \;\max_{p_{X^K U^K W|S} \in \scrP_{X^K U^K W|S}(p_S,L_w,L_u,D_1)}
	\;\min_{p_{Y|X^K} \in \scrW_K}
	\min_{\calA \subseteq \calK} J_{L_w,L_u,\calA}(p_S,p_{X^K U^K W|S}, p_{Y|X^K})\nonumber\\
&\stackrel{(b)}{=}& \sup_{L_w,L_u}\; \overline{C}_{L_w,L_u}^{all}(D_1,\scrW_K) \nonumber\\
&=& \lim_{L_w,L_u \to\infty} \overline{C}_{L_w,L_u}^{all}(D_1,\scrW_K) \nonumber\\
&\stackrel{(c)}{=}& \overline{C}^{all}(D_1,\scrW_K),
\label{eq:Converse C}
\end{eqnarray}
where
(a) holds because the functionals $J_{L_w,L_u,\calA}(\cdot)$ are nondecreasing in $L_w,L_u$,
(b) uses the definition of $\overline{C}_{L_w,L_u}^{all}$ in (\ref{eq:CL-all}), and
(c) the fact that the sequence $\{\overline{C}_{L_w,L_u}^{all}\}$ is nondecreasing.

\newpage

{\bf Proof of (\ref{eq:I-GPineq-cond})}.
Recall the definitions of $V_{\calK,i} = (M_{\calK}, V, S_{i+1}^N)$ and
$U_{\calK,i} = (V_{\calK,i}, (YS^d)^{i-1})$ in (\ref{eq:VW})
and the recursion (\ref{eq:VW-recursive}) for $V_{\calK,i}$.
We prove the following inequality:
\begin{eqnarray}
   \lefteqn{I(U_{\calA,i};Y_i S_i^d|U_{\calK\setminus\calA,i}) 
	- I(U_{\calA,i};S_i|U_{\calK\setminus\calA,i})} \nonumber \\
		& = & [I(V_{\calA,i};(YS^d)^i |V_{\calK\setminus\calA,i})
			- I(V_{\calA,i};S^i |V_{\calK\setminus\calA,i})] \nonumber \\ 
		& & - [I(V_{\calA,i-1};(YS^d)^{i-1} |V_{\calK\setminus\calA,i-1})
			- I(V_{\calA,i-1};S^{i-1} |V_{\calK\setminus\calA,i-1})] .
\label{eq:GP-recursion}
\end{eqnarray}
Then summing both sides of this equality from $i=2$ to $N$, cancelling terms, and using
the properties $V_{k,1} = U_{k,1}$ and $V_{k,N} = (M_k, V)$ yields (\ref{eq:I-GPineq-cond}).

The first of the six terms in (\ref{eq:GP-recursion})
may be expanded as follows:
\begin{eqnarray}
   I(U_{\calA,i};Y_i S_i^d|U_{\calK\setminus\calA,i})
	& = & I(V_{\calA,i},(YS^d)^{i-1};Y_i S_i^d|V_{\calK\setminus\calA,i},(YS^d)^{i-1}) \nonumber \\
	& = & I(V_{\calA,i};Y_i S_i^d|V_{\calK\setminus\calA,i},(YS^d)^{i-1}) \nonumber \\
	& = & I(V_{\calA,i},(YS^d)^{i-1};Y_i S_i^d|V_{\calK\setminus\calA,i})
			- I((YS^d)^{i-1};Y_i S_i^d|V_{\calK\setminus\calA,i}) \nonumber \\
	& = & I(U_{\calA,i};Y_i S_i^d|V_{\calK\setminus\calA,i})
			- I((YS^d)^{i-1};Y_i S_i^d|V_{\calK\setminus\calA,i}) .
\label{eq:GP-a}
\end{eqnarray}
Similarly for the second term, replacing $(YS^d)$ with $S$ in the above derivation,
we obtain
\begin{equation}
  I(U_{\calA,i};S_i|U_{\calK\setminus\calA,i})
	= I(U_{\calA,i};S_i|V_{\calK\setminus\calA,i})
			- I((YS^d)^{i-1};S_i|V_{\calK\setminus\calA,i}) .
\label{eq:GP-b}
\end{equation}
The six terms in (\ref{eq:GP-recursion}) can be expanded using the chain rule
for mutual information, in the same way as in \cite[Lemma~4.2]{Gelfand80}:
\begin{eqnarray}
   I(V_{\calA,i};(YS^d)^i |V_{\calK\setminus\calA,i})
	& = & I(V_{\calA,i};(YS^d)^{i-1} |V_{\calK\setminus\calA,i})
		+ I(V_{\calA,i};(YS^d)_i |V_{\calK\setminus\calA,i}) \label{eq:GP-1} \\
   I(V_{\calA,i};S^i |V_{\calK\setminus\calA,i})
	& = & I(V_{\calA,i};S^{i-1} |V_{\calK\setminus\calA,i})
		+ I(V_{\calA,i};S_i |V_{\calK\setminus\calA,i}) \label{eq:GP-2} \\
   I(V_{\calA,i-1};S^{i-1} |V_{\calK\setminus\calA,i-1})
	& = & I(V_{\calA,i};S^{i-1} |S_i, V_{\calK\setminus\calA,i-1}) \label{eq:GP-3} \\
   I(V_{\calA,i-1};(YS^d)^{i-1} |V_{\calK\setminus\calA,i-1})
	& = & I(V_{\calA,i};(YS^d)^{i-1} |S_i, V_{\calK\setminus\calA,i-1}) \label{eq:GP-4} \\
   I(U_{\calA,i};S_i |V_{\calK\setminus\calA,i})
	& = & I((YS^d)^{i-1};S_i |V_{\calK\setminus\calA,i})
		+ I(V_{\calA,i};S_i |(YS^d)^{i-1}, V_{\calK\setminus\calA,i}) \label{eq:GP-5} \\
   I(U_{\calA,i};(YS^d)_i |V_{\calK\setminus\calA,i})
	& = & I((YS^d)^{i-1};(YS^d)_i |V_{\calK\setminus\calA,i})
		+ I(V_{\calA,i};(YS^d)_i |(YS^d)^{i-1}, V_{\calK\setminus\calA,i}) . \nonumber \\
																		\label{eq:GP-6}
\end{eqnarray}
Moreover, expanding the conditional mutual information
$I(V_{\calA,i};S_i,(YS^d)^{i-1} |V_{\calK\setminus\calA,i})$
in two different ways, we obtain
\begin{eqnarray}
   \lefteqn{I(V_{\calA,i};(YS^d)^{i-1} |V_{\calK\setminus\calA,i})
		+ I(V_{\calA,i};S_i|(YS^d)^{i-1}, V_{\calK\setminus\calA,i})} \nonumber \\
	& = & I(V_{\calA,i};S^{i-1} |V_{\calK\setminus\calA,i})
		+ I(V_{\calA,i};(YS^d)^{i-1} |S_i, V_{\calK\setminus\calA,i}) .
\label{eq:GP-7}
\end{eqnarray}
Substracting the sum of (\ref{eq:GP-b}), (\ref{eq:GP-1}), (\ref{eq:GP-3}), (\ref{eq:GP-5}),
(\ref{eq:GP-7}) from the sum of (\ref{eq:GP-a}), (\ref{eq:GP-2}), (\ref{eq:GP-4}), (\ref{eq:GP-6}),
and cancelling terms, we obtain (\ref{eq:GP-recursion}), from which the claim follows.
\hfill $\Box$





\end{document}